\documentclass[%
 reprint,
 amsmath,amssymb,
 aps,
]{revtex4-2}

\usepackage[utf8x]{inputenc}
\usepackage{amsmath}
\usepackage{float}

\usepackage{subfigure} % subfiguras
\usepackage{graphicx}% Include figure files

\usepackage{dcolumn}% Align table columns on decimal point
\usepackage{bm}% bold math
\usepackage{hyperref}% add hypertext capabilities
\usepackage[dvipsnames,table,xcdraw]{xcolor}
\usepackage{changepage}
\usepackage{textcomp,marvosym}

\usepackage{nameref,hyperref}

\usepackage{microtype}
\DisableLigatures[f]{encoding = *, family = * }

\usepackage[table]{xcolor}

\usepackage{array}

\newcolumntype{+}{!{\vrule width 2pt}}

\newlength\savedwidth

\newcommand\thickhline{\noalign{\global\savedwidth\arrayrulewidth\global\arrayrulewidth 2pt}%
\hline
\noalign{\global\arrayrulewidth\savedwidth}}

\begin{document}

\preprint{APS/123-QED}

\title{Meteorological indicators of dengue epidemics in non-endemic Northwest Argentina}% Force line breaks with \\ 
\author{Javier Armando Gutierrez}
\altaffiliation[1]{Departamento de Física - FCE - Universidad Nacional de Salta, Salta, Argentina.}
\email{g.javier.ar@gmail.com}
\author{Karina Laneri}
\altaffiliation{Grupo de F\'isica Estad\'istica e Interdisciplinaria, CONICET, Centro At{\'{o}}mico Bariloche-CNEA, San Carlos de Bariloche, R\'{\i}o Negro, Argentina.}
\author{Juan Pablo Aparicio}
\altaffiliation{Instituto de Investigaciones en Energ\'ia no Convencional (INENCO)   Consejo Nacional de Investigaciones Cient\'ificas y T\'ecnicas (CONICET) - Universidad Nacional de Salta, Av. Bolivia 5100, 4400 Salta, Argentina.}
\altaffiliation{Simon A. Levin Mathematical, Computational and Modeling Sciences Center Arizona State University, PO Box 871904 Tempe, AZ 85287-1904, USA.}

\author{Gustavo Javier Sibona}
\altaffiliation{ IFEG - CONICET and Fa.M.A.F. - Universidad Nacional de C\'ordoba, C\'ordoba, Argentina.}

\date{\today}% It is always \today, today,
             %  but any date may be explicitly specified
%
\begin{abstract}
\hfill \break % en lugar de utilizar \\ 

In the last two decades dengue cases increased significantly throughout the world. In several regions dengue re-emerged, particularly in Latin America, where dengue cases not only increased but also occurred more frequently. It is therefore necessary to understand the mechanisms that drive epidemic outbreaks in non-endemic regions, to help in the design of control strategies.
We develop a stochastic model that includes climate variables, social structure, and mobility between a non-endemic city and an endemic area.
We choose as a case study the non-endemic city of San Ram{\'o}n de la Nueva Or{\'a}n, located in Northwest Argentina. Human mobility is intense through the border with Bolivia, where dengue transmission is sustained during the whole year. City population was modelled as a meta-population taking into account households and population data for each patch. Climate variability was considered by including rainfall, relative humidity and temperature time series into the models. Those climatic variables were input of a mosquito population ecological model, which in turn is coupled to an epidemiological model.
Different hypotheses regarding people's mobility between an endemic and non-endemic area are tested, taking into account the local climatic variation, typical of the non-endemic city. Simulations are qualitatively consistent with weekly clinical data reported from 2009 to 2016. Our model results allow to explain the observed pattern of outbreaks, that alternates large dengue epidemics and several years with smaller outbreaks. We found that the number of vectors per host and an effective reproductive number are proxies for large epidemics, both related with climate variability such as rainfall and temperature, opening the possibility to test these meteorological variables for forecast purposes.

\end{abstract}
%Use showkeys class option if keyword
                              %display desired
\maketitle
%\tableofcontents
%
\section{\label{sec:level1}Introduction}
\label{intro}

Several vector-borne diseases affect tropical, subtropical and also temperate regions worldwide \cite{Robert2019}. Dengue in particular, causes 390 million infections in the world every year, with nearly 3.9 billion people at risk of infection \cite{WHO}. Reported dengue cases increased significantly in some Latin American countries in the last 20 years \cite{WHO}. Four different antigenically related serotypes, i.e. DENV-1, DENV-2, DENV-3 and DENV-4, produce a large variety of symptoms, from mild to severe and rarely fatal ones. Although each virus strain provides partial temporal immunity for that specific serotype, there is no cross-immunity. Antibodies generated by previous infections cross react with infections of a different strain, sometimes increasing symptoms severity \cite{WHO}.

Dengue virus transmission occurs in a mosquito-human cycle. The vectors that transmit dengue disease in the Latin American region are \textit{Aedes albopictus} and \textit{Aedes aegypti}, being the last one more prevalent throughout the region \cite{kraemeretal,TORRES2007,Lopez-Gatell2015}. \textit{Aedes aegypti} has domestic habits and mainly breeds on artificial water containers \cite{Harrington1998}. Therefore climate variability might not be a good candidate to drive dengue cases variability. However some studies show that vector survival and virus development are constrained, specially by rainfall and temperature \cite{romeoaznar2018}, and others are successful in predicting dengue incidence with climate-driven models \cite{vasquez2019}. Therefore, to which extent the variability of some climatic variables can anticipate the timing and relative intensity of dengue outbreaks in our studied region is still unknown.

Dengue is endemic in many countries, i.e. transmission is present during the whole year. By contrast, in non-endemic regions, like Northwest Argentina \cite{aviles1999,rotela2007}, dengue outbreaks occur during a given season. A typical case with both regions is Taiwan, where the south is endemic while the north of the island is epidemic and seasonal \cite{valdezLD2018}. In Latin America both endemic and epidemic/seasonal dengue dynamics are observed. 
For the particular case of Northwest Argentina, outbreaks occur during the hot rainy season (November-March). Moreover, during most years there are few cases but large epidemics were observed in 2004, 2009 and 2016 with thousands of cases and several deaths, with a significant economic impact in the population \cite{tarragona2012}. Therefore, forecasting the risk of a large epidemic in these non-endemic regions may help to take preventive actions to reduce negative impacts of the disease.

Previous works developed different models \cite{kramer2015,REINER2014,SARDAR2015} to describe the dynamics of dengue transmission. 
We linked a stochastic simulation model for the \textit{Aedes aegypti} population dynamics \cite{valdezLD2018} to an epidemiological model to explore the impact of imported cases flux and climate variability on dengue disease dynamics. We show how the interaction of all these factors influence the variability of dengue cases recorded between 2009 and 2016 into a non-endemic city of Northwest Argentina.

\section{Materials and methods}

\begin{figure}[ht]
\centering
\includegraphics[width = 8.5 cm]{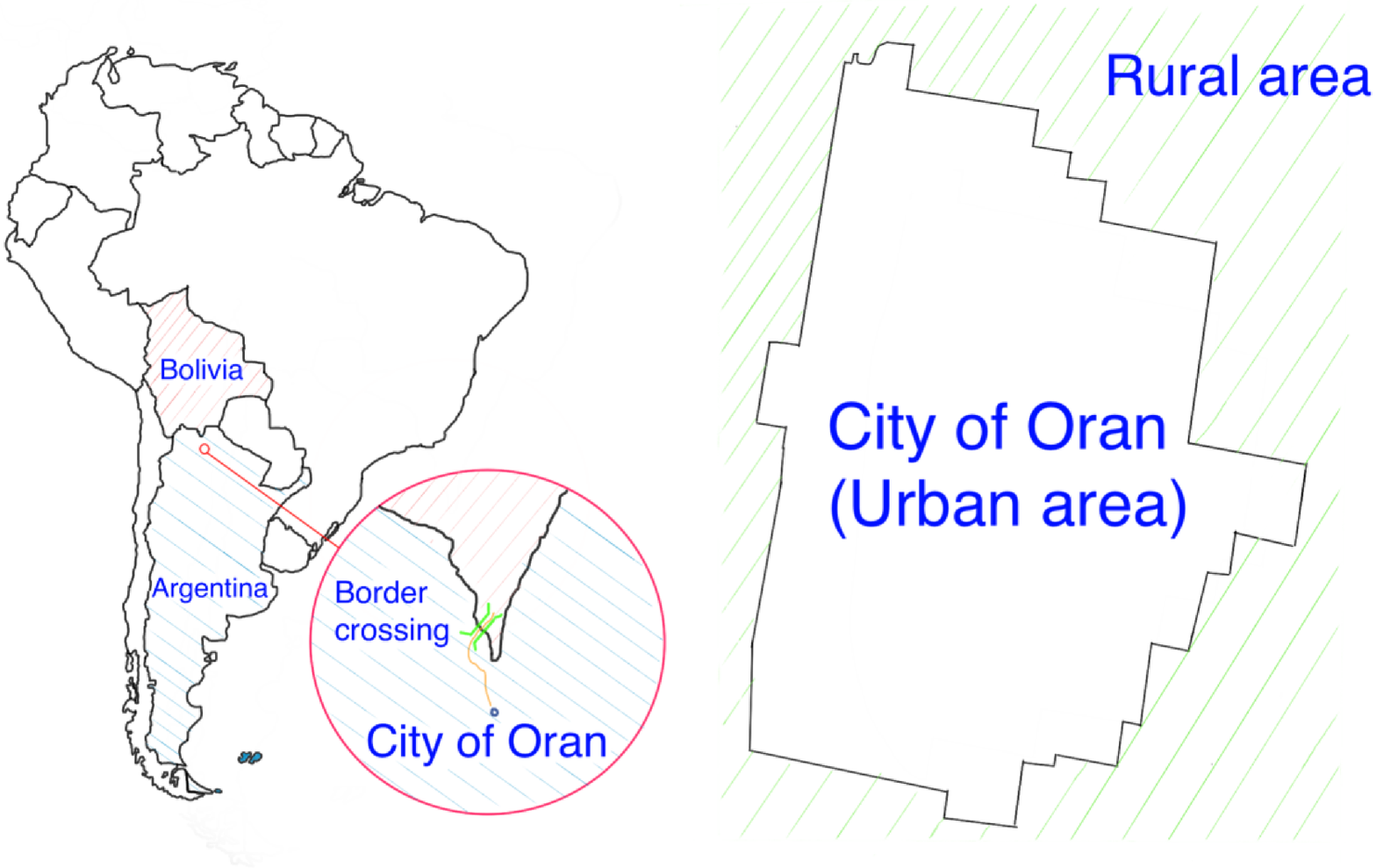}
\caption{\textbf{Geographical location of the city of Oran}. From left to right: map of South America, the zoomed region shows the Argentina-Bolivia border, sketch of the city of Oran. The map was done by the authors under the common creative license.}
\label{mapa_oran}
\end{figure}

We will take as a case study the city of San Ram{\'o}n de la Nueva Or{\'a}n (Oran from now on), although the model is applicable to any non-endemic city. This city has an extension of $3.75$ km $\times$ 5 km, a population of 75697 people (17633 houses), and is located in Northwest Argentina (${23}^\circ 07'45.8"S$, ${64}^\circ 19'16.9"W$, see Fig \ref{mapa_oran}). It is 50 km south of the Bolivian-Argentinian border, where is connected with Bermejo city (Bolivia) through a bridge that crosses Bermejo river. Therefore social and business activities across the border are very intense.

\begin{figure}[ht]
\centering
\includegraphics[width = 8.5 cm]{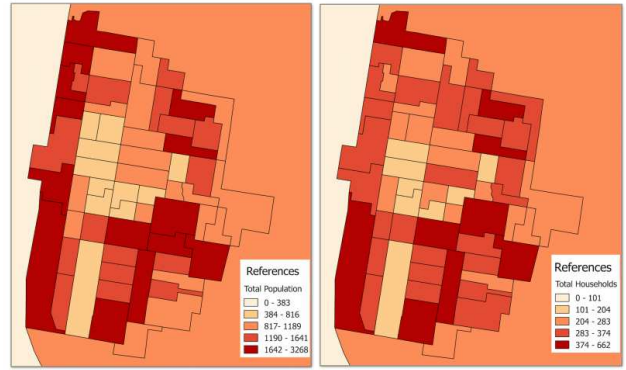}
\caption{\textbf{Distribution of populations and households in the city of Oran.} Jenks categorization for the population (left) and households (right) of the city of Oran. In this work we do not consider the two largest rural ratios (rural areas) \cite{INDEC}}
\label{metapop}
\end{figure}

In 2010, a population census was conducted in Argentina \cite{INDEC}. For census purposes the city was partitioned in polygons or census sections, each of which will be considered as a patch for our meta-population model. From the census data \cite{INDEC} we obtained population size and number of houses for each patch (Fig \ref{metapop}). The city of Oran contains a total of sixty urban sections and two rural sections. Surrounding rural areas, including the two rural sections, are not considered in our model.

%\subsection{Climate}
Oran has a tropical climate; with dry, mild winters and hot summers with high rainfall (see Section S4 in Supplementary Material). As a consequence, the population of \textit{Aedes aegypti}, the main dengue transmission vector in the region, shows marked seasonal variations \cite{estallo2011}.
In fact, the appearance of new cases of dengue is related in part with an increase in the vector population. 

We developed a stochastic meta-population model for disease transmission and mosquito population dynamics. For each patch in Oran, we considered its human and vector population size, possible imported cases that arrive, the number of households, as well as an estimated number of mosquito breeding sites. 
Imported cases were recorded by the National Ministry of Health of Argentina (see Data Statement). 

Given that {\it Aedes} mosquitoes usually breed in domestic containers, we assumed as a modeling hypotheses, that breeding sites were proportional to human population size that was estimated as the number of houses in each patch. Vector population size, as well as number of mosquitoes and breeding sites, were estimated by performing computer simulations with the mathematical model described in the section below (see "Ecological model for {\it Aedes aegypti}").

Daily time series of relative humidity (RH), temperature (T) and precipitation for the period 2009–2017, were provided by the Argentinian Meteorological Service. Daily averages of those climatic variables were used as input for the ecological model for {\it Aedes aegypti} population, although qualitatively similar results were obtained when using their maximum or minimum values.

For the simulations we used the Poisson approximation \cite{aparicio2001a} with a fixed time step $\Delta t$. For each event considered in the model we computed the corresponding transition rate, i.e. the probability of occurrence per unit of time, generating a pseudo random number following a Poisson distribution with parameter equal to the transition rate, times $\Delta t$ (see S7 in Supplementary Material). 

\begin{figure}
\centering
\includegraphics[width = 8.5 cm]{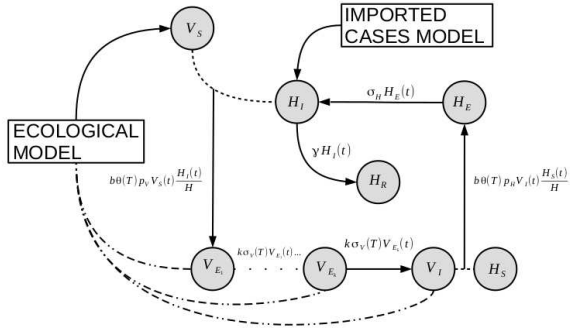}
\caption{\textbf{Scheme of the epidemiological model.} $H_S,H_E,H_I,H_R$ are susceptible, exposed, infected and recovered hosts populations respectively. Similarly $V_S,V_{E_i},V_I$ are susceptible, exposed and infected vector populations respectively. Sub-index $i$ refers to different mosquito stages in exposed class.}
\label{modelo_epi}
\end{figure}

%%%%%%%%%%%%%%%%%%%%%%%%%%%%%%%%%%%%%%%%%%%%%%%%%%%%%%%%%%%%%
Demographic variables do not change in the studied time period, therefore we considered for each patch in the city a constant human host population of size $H$ divided in four epidemiological classes: susceptible ($H_S$), exposed ($H_E$), infectious ($H_I$) and recovered ($H_R$). As mosquitoes are infected for life we considered only susceptible ($V_S$), exposed ($V_E$) and infectious ($V_I$) vector populations. A scheme of the epidemic cycle is depicted in Fig \ref{modelo_epi}.

The complete model implemented to perform the computer simulations consists in a compartmental model (see Figure \ref{modelo_epi}) in combination with an ecological model for the mosquitoes population and an imported cases model, together with the vector-host model. These models are applied on a meta-population structure that emulates a no-endemic city. For each patch, information such as the number of hosts and households is included, which allows to reproduce the real city heterogeneities. In the following sections we briefly describe those model components.

\subsection*{Epidemiological model}

Virus transmission was modelled using frequency dependent rates. The susceptible mosquitoes get infected while in contact with an infected host. Thus the probability of vector infection per unit of time ($G_V(t;T)$) at a given temperature $T$, is given by:
\begin{equation}
G_V(t;T) =  b  \theta(T) p_V V_S(t)  \frac{H_I (t) }{H}   \nonumber
\end{equation}
where $b$ is the mosquito biting rate, $\theta(T)$ is the effect of temperature on the biting rate and $p_V$ is the probability of infection per bite for vectors. A similar probability could be written for human infection:
\begin{equation}
G_H(t;T) = b  \theta(T) p_H   V_I (t)  \frac{H_S (t)}{H}     \nonumber
\end{equation}
with $p_H$ the infection probability per bite for hosts.

To compute the latency period of vectors before becoming infectious we follow the model proposed in \cite{romeoaznar2015} which considers a series of steps, exponentially distributed in time, performed in sequence to complete an infectious process. The convolution of these steps leads to a gamma distribution for the latency period (see Supplementary Material \ref{latency_aedes}). 

For humans, the progression from latency to the infectious state takes place at a constant rate $\sigma_H$, while recovery rate $\gamma$ is also constant. In other words, we assume that latency and infectious periods are exponentially distributed with mean values $1/\sigma_H$ and $1/\gamma$ respectively. 

Dispersion of \textit{Aedes aegypti} is estimated in a few hundred meters \cite{sowilem2013}, while the size of the patch is of the order of 800m, therefore we only considered mosquito dispersion to neighbouring patches using a dispersion model \cite{javier2015} with low diffusion values ($D_V < 0.1$ and $D_V = 0$). Simulations performed with both diffusion values show no significant variations. Therefore to save computational costs we choose $D_V = 0$, avoiding the need to make the balance of agents (vectors and hosts) in each time step.
The main source for disease spreading between patches is human mobility. For the spatial dynamics of the human population we will consider that individuals may visit any randomly chosen patch. In this way, it is possible the transmission of the virus from a visiting infected host to susceptible vectors in another patch, or the transmission to a susceptible host visiting a patch with infected vectors. To model this dynamics we included a term considering the possibility of disease transmission between vector and host of neighbouring patches or for a randomly chosen pair of patches at each time step. Under these assumptions the probability of vector infection per unit of time in patch $j$ is:

\begin{equation}
\label{local_vector}
\mathrm{Inc}_V^j=b  \theta(T)  p_V   \frac{P_{H_I}(t)}{P_H} {V_S(t)}^j \ 
\end{equation}
where $P_{H_I}$ and $P_H$ are the sum of infected hosts and the total number of hosts in the patch, in its nearest neighbours, and in a randomly chosen patch in the time step. 
The number of new infected vectors was simulated by drawing a random number, Poisson distributed, with mean $\mathrm{Inc}_V^j  \Delta t$.

In an analogous way, the number of new infected hosts in a time step was simulated using the probability of human infection per unit of time given by:

\begin{equation}
\label{local_host}
\mathrm{Inc}_H^j = b \theta(T)   p_H   \frac{ {H_S(t)}^j }{P_H}  P_{V_I(t)}
\end{equation}
where $P_{V_I}$ is the sum of infected vectors in patch $j$, in its nearest neighbours, and in a patch randomly chosen at each time step.

\subsection*{Model for imported cases} \label{imported_cases}
In our sub-tropical non-endemic region, dengue  transmission is interrupted during the coolest dry season. Therefore, assuming a negligible vertical transmission for {\it Aedes aegypti}, new outbreaks mostly start by the influx of infected people. 
These imported cases, i.e. individuals who contracted the virus in endemic areas, can be people from Argentina returning from holidays in Bolivia, or people doing temporary travel to Argentina for commercial or working activities.

The flux of imported dengue cases is modelled as:
\begin{equation}
\label{equation_imported}
\lambda_I(t) = p_i  F_ {\mathrm {PS}} (t)  \frac{I_B(t)}{N_B}
\end{equation}
where $F_ {\mathrm {PS}}(t)$ is the number of people from the non-endemic area who are in contact with people from the endemic area, $\frac{I_B(t)}{N_B}$ is the prevalence of infectiousness in the endemic area and $p_i$ is the rate of new imported cases per host. From Eq. \ref{equation_imported} we constructed the time series of imported cases $\delta_I(t)$, sampling random numbers from a Poisson distribution with parameter $\lambda_I(t) \Delta{t}$. We tested the model for several $F_ {\mathrm {PS}}(t)$ representing the different imported cases scenarios considered.

Each imported case was assigned to a randomly chosen patch. See \ref{serie_import} of the Supplementary Material for more details.

\subsection*{Ecological model for \textit{Aedes aegypti}}
\label{ecological_model}

\begin{figure}[ht!]
\centering
\includegraphics[width = 8 cm]{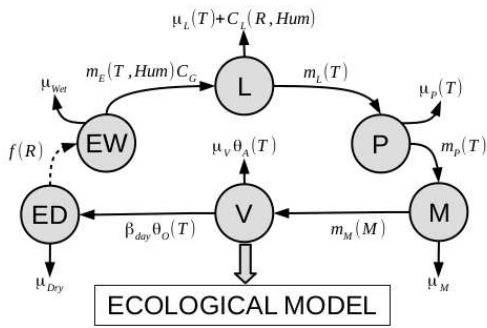}
\caption{\textbf{Scheme of the ecological model for vector population.} Circles represent the different stages of \textit{Aedes aegypti} maturation: dried egg ($E_D$), wet eggs ($E_W$), larvae ($L$), pupae ($P$), young mosquitoes ($M$) and adult female mosquitoes ($V$). Solid arrows indicate those processes that occur at a constant rate, while the dotted arrow indicates a fixed period process that triggers depending on the rain. }
\label{ecological}
\end{figure}

The mosquito population dynamic model (Fig \ref{ecological}) is inspired on the abundance model \cite{valdezLD2018}. Local weather conditions as temperature ($T$), precipitation level ($R$) and relative humidity (RH) strongly affect mosquito activity, development, and egg hatching.

We considered five stages for the mosquito population: eggs, larvae, pupae, pre-adults (adult mosquitoes with non fully developed wings) and adults. Eggs may be in two states, dry and wet, with population sizes $E_D$ and $E_W$ respectively. According to our model the larval capacity of each patch depends on the number of households, and on the climatic conditions. 

Recruitment rate of new adult susceptible mosquitoes, $\Lambda (T,R,RH)$,  depends on temperature ($T$), rainfall ($R$) and relative humidity ($RH$) and it is obtained from an ecological model for the mosquito population dynamics described in \ref{ecological_model_app} of the Supplementary Material.
Mosquito mortality, $\mu_V(T)$, is temperature dependent and we assumed that it does not depend on the mosquito age. This is a reasonable assumption given that the consideration of mosquito age classes, would significantly increase model complexity and computational cost without need, because mosquito lifetime is only limited to the summer season.

The complete model, that integrates all the processes detailed in the previous sections, was implemented and its outcome together with field data are presented in the following section.

\section{Results}

\begin{figure}[ht]
\centering
\includegraphics[width = 8.5 cm]{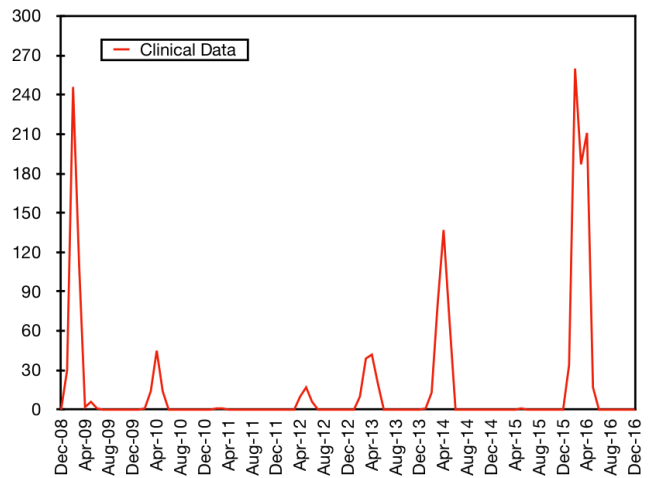}
\caption{\textbf{Dengue cases in the city of Oran.} Dengue monthly cases from December 2008 to December 2016. Positive cases are those confirmed by laboratory (viral confirmation and serotype identification, positive screening test). Epidemic outbreaks do not occur during the dry months (winter). Epidemics were mostly of DENV-1 in 2009 and of DENV-4 type in 2016. }
\label{casos_oran}
\end{figure}

Dengue re-emerged in Argentina in 1997. In the city of Oran
there were large epidemics beginning in 2009 and 2016, with small outbreaks in between, as shown by the time series of laboratory confirmed dengue cases (Fig \ref{casos_oran}).

Each year, the first dengue cases are imported cases, mostly due to infection of city residents who travelled to endemic areas of Bolivia, as reported for example in \cite{gil2016}. Cases detected by the Argentinian Ministry of Health are classified as imported or locally transmitted (see Data Statement). 

Usually, single-strain outbreaks are observed and, only in a few occasions, more than one strain circulates, being only one of them responsible of more than $90\%$ of the reported cases. Therefore, computer simulations of the epidemiological model developed in the present work were performed considering only one virus strain. 

In the following sections we present the results obtained from the combination of data and computer simulations with the complete model described in the previous section.

\subsection*{Meteorological variability and dengue dynamics}
\begin{figure}[ht!]
\centering
\includegraphics[width = 8.5 cm]{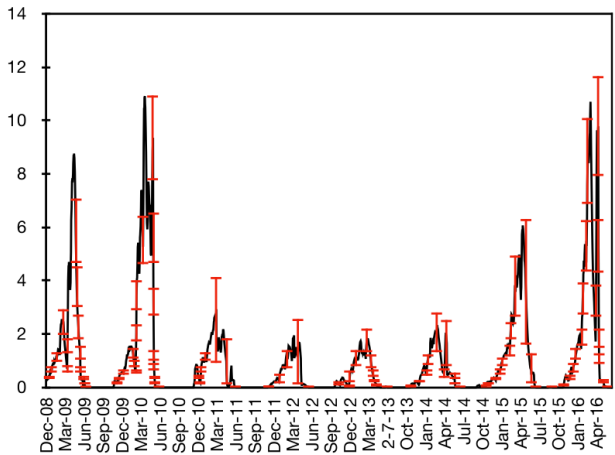}
\caption{\textbf{V/H Factor.} Representative simulation of the average vector-host ratio ($V(t)/H$) obtained with 100 simulations, with its standard error, summing up all the patches in Oran city.}
\label{vector_host}
\end{figure}

We simulated the evolution of the vector population with the ecological model for mosquito population dynamics, using meteorological time series as model input. The human population in each patch was estimated from census data. After computing the vector-host ratio for each patch, we computed its average over all the simulations for each of the patches. All the obtained values were summed up for each time step to obtain the $V(t)/H$ time series. Our model predicts two significantly larger values for the vector-host ratio $V/H$ for years 2009 and 2016 (Fig.~\ref{vector_host}). 

Those years were the ones with the largest dengue epidemics Fig.~\ref{casos_oran}. Therefore the average vector-host ratio could be an indicator of a high probability of epidemics in these particular years.

Given that the vector population dynamics obtained from the ecological model mainly depends on meteorological variability, the evolution of the number of vectors per host ($V(t)/H$) shows little variation between different simulations. Mosquito population is different among patches as it depends on the number of houses in each patch. However during simulations, the disease spread produced by infected mosquitoes between patches is negligible compared with the long range movement of hosts. Additionally, mosquito population inside each patch did not change significantly, due to low mosquito dispersal between patches given that patch area is larger than the lifetime dispersal area of \textit{Aedes aegypti} \cite{sowilem2013}. Therefore, we can conclude that meteorological factors constitute the main source of dengue cases variability.

\subsection*{Basic and effective reproduction numbers}

To model the dynamics of the mosquito population we considered $k$ gamma distributed latency periods. Neglecting the spatial effects (diffusion from and to other regions), the basic reproduction number ($R_0$) of our model is: 

\begin{equation}
\label{R_0}
R_0 = b^2  \theta(T) ^2  \frac{p_H p_V }{\mu_V(T)  \gamma}  {\left( \frac{k \sigma_V(T)  }{\mu_V(T) +  k  \sigma_V(T) } \right)}^k  \frac{V(t)}{H}
\end{equation}
Where ${\left( \frac{k \sigma_V(T)}{\mu_V(T) + k \sigma_V(T) } \right)}^k$ is the fraction of mosquitoes that survive to the different $k$ latency stages. The basic reproduction number is proportional to the number of vector per human, $V/H$. Details of the derivation of Eq. \ref{R_0} can be found in the Supplementary Material.

It is important to remark that we are interested in reproducing the
observed pattern, with large epidemics and small outbreaks in between, and not the number of reported cases, because in many cases only a fraction of the suspected cases are sent for laboratory confirmation.

As we are interested in recurrent outbreaks dynamics, a more useful measure of epidemic probability is the effective reproduction number ($R_0^*$) which takes into account that only a fraction of the host population is susceptible. Without considering spatial effects the effective reproduction number is: 
\begin{equation}
\label{R_0_eff}
R_0^* (t) = b^2  \theta(T) ^2 \frac{p_H p_V }{\mu_V(T)  \gamma}  {\left( \frac{ k  \sigma_V(T) }{\mu_V(T) +  k  \sigma_V(T) } \right)}^k  \frac{V(t)}{H} \frac{H_S(t)}{H}
\end{equation}
The extra factor $\frac{H_S(t)}{H}$ accounts for the amount of susceptible humans available in the system. As vector population is constantly renewing itself, for practical purposes we considered a constant proportion of susceptible mosquito population ($\frac{V_S(t)}{V}$) in Eq. (\ref{R_0_eff}).

\begin{table*}[!ht]
%\begin{adjustwidth}{-2.25in}{0in} % Comment out/remove adjustwidth environment if table fits in text column.

\centering
\caption{
{\bf Estimated values for the epidemiological model.}}
\begin{tabular}{|l|l|l|}
\hline
\multicolumn{1}{|l|}{\bf Parameter} & \multicolumn{1}{|l|}{\bf Value}  & \multicolumn{1}{|l|}{\bf Reference}\\ \thickhline
$\beta_{\mathrm{day}}$ & 2.39 (1.9 - 2.4)  [egg day$^{-1}$ ] & \cite{grech2010} and \ref{aedes_ovi} \\ \hline
$\mu_{Dry}$ & $120^{-1}$ [day$^{-1}$ ] & \cite{valdezLD2018} \\ \hline
$\mu_{Wet}$ & $90^{-1}$ [day$^{-1}$ ] & \cite{valdezLD2018} \\ \hline
$R_{\mathrm{thres}}$ & 10.5 (7.5 - 12.5) [mm] & \cite{valdezLD2018} \\ \hline
$H_{\mathrm{max}}$ & 24 [mm] & \cite{valdezLD2018} \\ \hline
$k_{\mathrm{I}}$ & $3.9 \times 10^{-5}$ ($3.3 \times 10^{-6},6.6 \times 10^{-5}$) [mm $^{o}$C$^2$] & \cite{valdezLD2018} \\ \hline
$K_{\mathrm{max}}$ & 120 (70 - 220) & \cite{valdezLD2018,estallo2011} \\ \hline
$\mu_V(T)$ & $10^{-1}$ $\eta(T)$ [day$^{-1}$] & \ref{mortality_aedes} \\  \hline
$\mu_M$ & 0.5 (0.25 - 0.5) [day$^{-1}$] & \\ \hline
$m_M$ & 0.5 (0.25 - 1) [day$^{-1}$] & \\ \hline 
$\sigma_V(T)$ & $ k \zeta(T)$  [day$^{-1}$] & \ref{latency_aedes} \\ \hline
$k$  & 4 (3 - 7) & \ref{latency_aedes} \\  \hline
$D_V$ & 0.0 (0  - 0.1) & \\  \hline
$b$  & 0.30 [day$^{-1}$ ] & \cite{dengueanuj}  \\ \hline
$p_V$ & 0.75 & \cite{wearing2006} \\ \hline
$p_H$ & 0.75 & \cite{wearing2006} \\ \hline
$\sigma_H$ & $4^{-1}$ [day$^{-1}$ ]  & \cite{wearing2006} \\ \hline
$\gamma$ &  $7^{-1}$ [day$^{-1}$ ] & \cite{wearing2006} \\ \hline
\end{tabular}
\begin{flushleft}  For the maturation rates see \ref{maturation_aedes}. In all simulations we estimate a $H_S(0) \approx 0.75$ for the city of Oran, each patch has initially 75\% population susceptible.
\end{flushleft}
\label{table1}
%\end{adjustwidth}
\end{table*}

\begin{figure}[ht!]
\centering
\includegraphics[width = 8.5 cm]{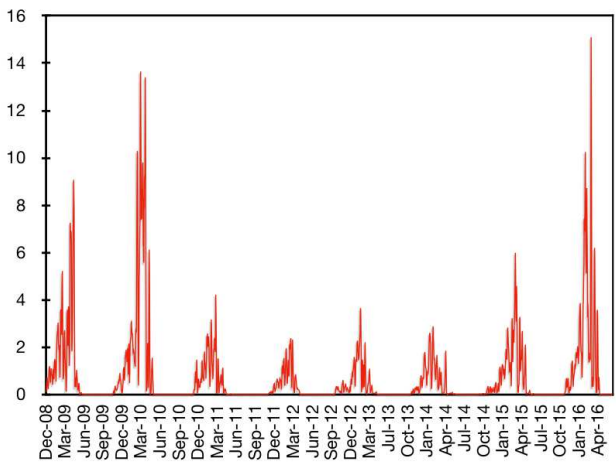}
\caption{\textbf{Basic effective reproductive number.} Representative graph of 100 simulation's moving average of period four days for the effective reproductive number.}
\label{r0_eff}
\end{figure}

 In this case we used the full model (see Eq. \ref{stoch_epidemiological_model}) to obtain estimations of the human susceptible population $H_S$. We computed the moving average of expression in Eq. \ref{R_0_eff} using a time window of four days (Fig. \ref{r0_eff}) to smooth the variability introduced by considering four infected mosquito latent stages. The pattern obtained for $R_0^* (t)$ is similar to the one of the fraction $V/H$ (Fig. \ref{vector_host}), with a large peak in 2016 besides the large peak of 2010. 
This is an interesting result despite the fact that a constant human population without recruitment of susceptibles is considered, and therefore the proportion of susceptible hosts necessarily decreases with successive outbreaks. Thus, due to meteorological variability is possible to obtain an epidemic after smaller outbreaks.
We can conclude that the effective reproduction number $R_0^* (t)$ (Eq. \ref{R_0_eff}) could be another indicator of dengue epidemics.

\subsection*{Influence of imported cases}

For Oran, the majority of imported cases comes from the nearby endemic area of Bolivia, as there is a fluid movement of people across the border.

To model the influx of imported cases from Bolivia we used data of dengue incidence in Bolivia \cite{PAHO} (see \ref{dengue_bolivia} in Supplementary Material) and estimations of people entering Salta from Bolivia (see \ref{serie_import} in Supplementary Material).

The influx of imported cases is a necessary source of new cases, as the disease cannot emerge every season in a non-endemic region without that input. Therefore as a first approach we assumed a number of imported cases proportional to the number of cases in Bolivia ($I_B(t)$ in Eq. \ref{equation_imported}). We proposed two different functional forms for the number of people entering from the endemic country ($F_{PS}(t)$ in Eq. \ref{equation_imported}), i.e. a constant influx during the whole year and a seasonal influx of imported cases. Averaged simulations do not reproduce the clinical cases time series. In other words, simulations assuming a fixed number of imported cases proportional to the number of cases in Bolivia, do not reproduce the observed time series, independently of how those imported cases are distributed throughout the year (See \ref{importedflux}, Figures \ref{imported_constant} and \ref{imported_seas} in Supplementary Material).

A more successful approach consisted in not imposing the functional shape for the flux of imported cases but instead in fitting $\lambda_I(t)$ (Eq. \ref{equation_imported}) for every year, to reproduce the time series of observed cases in Oran. This way, the number of imported cases emerges from the dataset and it is not necessarily proportional to the number of cases in Bolivia, which might be a reasonable assumption as we will discuss later.

\begin{figure}[ht!]

\centering
\subfigure{\includegraphics[width = 8 cm]{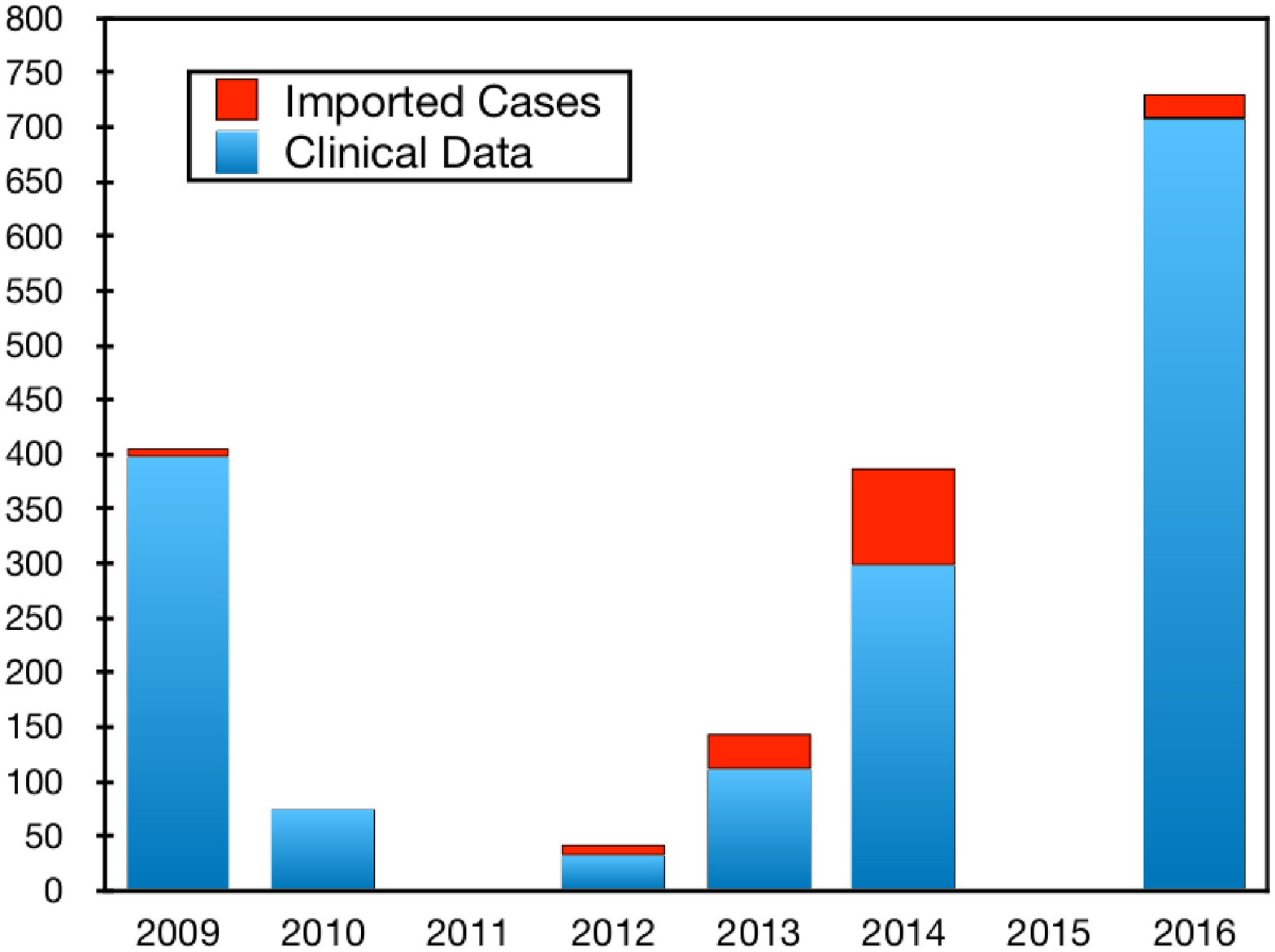}}
\subfigure{\includegraphics[width = 8 cm]{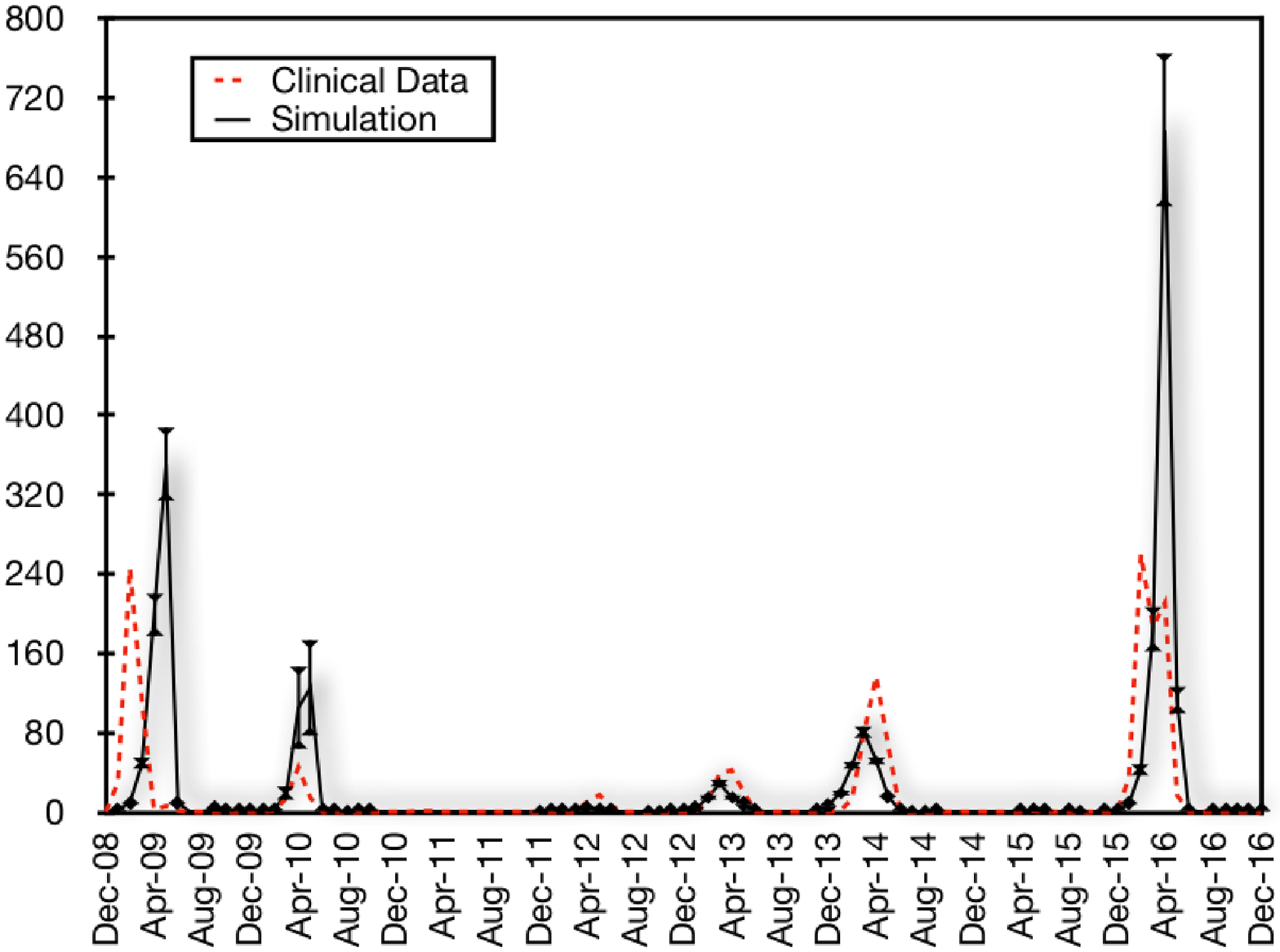}}

\caption{(a): bottom bars represent confirmed cases in Oran. Top bars show the number of imported cases from 2009 to 2016 after fitting to data in the bottom panel. The percentages of imported cases relative to the total number of cases resulted: 1.35\%, 50.00\%, 36.36\%, 29.73\%, 30.20\%, 100.00\% and 1.41\%. (b): comparison between observed and simulated dengue incidence using the fitted values for imported cases. The series is an average of 100 simulations and the bars correspond to the standard error.}
\label{importados_ajustados}
\end{figure}

To construct the $\delta_I(t)$ time series, imported cases were randomly distributed evenly over the first four months of each year that correspond to the rainy period, fitting on average the observations (as explained in previous sections). The procedure is repeated for each iteration. A new time series $\delta_I(t)$ is constructed, that correlates with the reported cases for each year. Using the parameters of \textbf{Table 1}, we compare simulations and clinical data (see Fig. \ref{importados_ajustados}A), assessing the values and the corresponding error of imported cases for each epidemic year (see Fig. \ref{importados_ajustados}B).

According to reports given by Argentina's epidemiological surveillance bulletin \cite{MSPN1,MSPN2}, imported dengue cases reported for 2016 represented no more than 10\% of the total cases in the province of Salta. Most dengue fever cases in that province occur in the city of Oran, since Tartagal, the other comparable city of the region, has dengue prevention programmes ongoing \cite{mundosano}. This percentage of imported cases reported in 2016, is in agreement with our estimation for that year (1.41\% as shown in Fig. \ref{importados_ajustados}).

We can therefore conclude that the estimated number of imported cases per year trigger the corresponding outbreak but does not determine its size. In other words, years with large epidemics might be triggered by a small number of imported cases. Favorable meteorological conditions for the increase of vector population are an essential determinant for large epidemic size.

\subsection*{Role of stochasticity and spatiality.}
Deterministic models always produce a large initial epidemic followed by substantially smaller outbreaks \cite{tesis2018}. Therefore, stochasticity is essential to reproduce the pattern of observed outbreaks, i.e. two epidemics and smaller outbreaks in between (see Fig. \ref{importados_ajustados}). 
In other words, because of the stochastic character of the system small outbreaks are common, and therefore the human susceptible population is almost not reduced, leaving enough susceptible individuals for future large outbreaks; something impossible to observe with deterministic models. However stochasticity alone is not enough. 

Spatiality provides a small and clustered number of vector-human contacts which is also a critical factor to reduce the size of small outbreaks. Upon a mosquito population threshold, the probability of large outbreaks increases and therefore epidemics have a very high probability of occurrence.

\section{Discussion}

Qualitatively comparing model and data, we found two proxies for epidemics that might be of importance. On the one hand the fraction of vectors by host, that is closely related with rainfall and temperature. This dependence is presumably related to the maturation of mosquito eggs available to hatch at the beginning of the season, which is closely related to the amount of rainfall. Additionally it is necessary to have an stable high temperature without significant low values during the season. For instance, it's interesting to notice that for epidemic years 2009 and 2016 the frequency of four consecutive days in a given range of temperature was higher than most years with small outbreaks.

On the other hand the effective reproductive number $R_0^*$, that is a product of the vector by host ratio and also a function of temperature, was found to be another epidemic proxy. The relevance of the link between vector abundance and human density for vector borne diseases was reported recently \cite{romeoaznar2018}; therefore our findings indicate that, for quantitative comparisons between model and data, the functional dependence between vectors and hosts should be further explored.

The number of imported cases emerging after fitting the dataset was not proportional to the number of dengue cases in Bolivia. This is a reasonable result given that other factors might influence that quantity, e.g. economic exchange between countries, commercial activities as well as temporal work might drive the variability found in the number of imported cases. When imported cases appear in a sustained manner, and with conditions as previously described (by climate and location), the probability of an epidemic outbreak is very high.

The results obtained are quite satisfactory in terms of their qualitative comparison with clinical data. A quantitative comparison became extremely difficult given that confirmed cases are not the true total number of cases, not only because a considerable number of cases might be asymptomatic but also due to delays in cases recording, as the health system is centralized in Buenos Aires and becomes overwhelmed in epidemic situations.

\section{Conclusion}
Climate therefore constitutes a fundamental component in the population dynamics of \textit{Aedes aegypti} and in the probability of dengue transmission. All these results open a promising door to explore the potential of meteorological data as a predictive tool for dengue epidemics. To address this formidable task is necessary to have access to good quality dengue cases records as well as to local rainfall and temperature forecasts. If these datasets were available, quantitative results would be obtained by fitting our model to data and performing intensive computer simulations with the ensemble of previously fitted parameters. Using climate forecasts as input it would be possible to obtain model hindcasts and, a posteriori, dengue outbreak forecasts (using for example V/H and $R_0^*$ as indicators). If the obtained uncertainties are acceptable, those forecasts can help to foresee new dengue outbreaks and alert the health system with enough anticipation to design strategies for dengue control. Given that Northwest Argentina is a hot spot for climate change, these relationships between climate and epidemics represent a valuable finding that should be deeply studied in the near future. 

\section*{Author Contributions}
JAG: Model development, collection of climatic and social data, programming, draft paper writing, corrections and discussion. KF: Collection of clinical data, corrections and discussion. JPA, GJS: Corrections and discussion.

\begin{acknowledgments}
This work was partially supported by grants PICT 2017-3117, PIP 112-2015 01-00644CO, SeCYT-UNC 05/B457 and PICT-2019-2019-03558. We thank the Ministry of Health of Argentina and the National Meteorological Service of Argentina for providing the datasets. KL, GJS and JPA are members of Consejo Nacional de Investigaciones Científicas y Técnicas (CONICET), Argentina.
\end{acknowledgments}

\bibliography{myreferences}% Produces the bibliography via BibTeX.

\clearpage
\appendix
\addcontentsline{toc}{part}{Appendix}%TOC entry without page numbering 

\renewcommand{\thesection}{S\arabic{section}}
\renewcommand{\thetable}{S\arabic{table}}
\renewcommand{\thefigure}{S\arabic{figure}}
\renewcommand{\theequation}{S\arabic{equation}}
\setcounter{equation}{0}
\setcounter{section}{0}
\setcounter{table}{0}
\setcounter{page}{1}
\setcounter{figure}{0}
%\documentclass[a4paper,10pt]{article}
%\usepackage[left=1.5cm,top=2cm,right=1.5cm,bottom=2cm]{geometry} 
%\usepackage[utf8]{inputenc}
%\usepackage{epsfig}
%\usepackage{capt -of}
%\usepackage{fancyhdr}

%package ulem for strike-through (Karina)
%\usepackage{ulem}
% color can be used to apply background shading to table cells only
%\usepackage[table]{xcolor}
%%%%%%%

%\pagestyle{plain} %%por defecto
%%sin encabezado y con numeracion
%\pagestyle{headings} %%con encabezado y numeracion
%\pagestyle{empty}

\newpage
\clearpage
\renewcommand{\thesection}{S\arabic{section}}
\renewcommand{\thetable}{S\arabic{table}}
\renewcommand{\thefigure}{S\arabic{figure}}
\renewcommand{\theequation}{S\arabic{equation}}
\setcounter{equation}{0}
\setcounter{section}{0}
\setcounter{table}{0}
\setcounter{page}{1}
\setcounter{figure}{0}

\pagestyle{myheadings}
\markright{Supplementary Material}

%opening

\title{Meteorological indicators of dengue epidemics in non-endemic Northwest Argentina}

\author{Supplementary Material}
\date{2021}
%\begin{document}
\maketitle

\section{Epidemiological Model}
\label{stoch_epidemiological_model}
Dengue transmission includes a vector-host cycle and can be modelled using a SEIR (Susceptible - Exposed - Infected - Recovered) model for hosts and an SEI (Susceptible - Exposed - Infected) model for vectors.

The epidemiological model developed in the present work consist in a compartmental model (see Fig.~\ref{modelo_epi}) in combination with an ecological model for the mosquitoes population and an imported cases model, together with the vector-host model. These models are applied on a meta-population structure that emulates a no-endemic city. For each patch, information such as the number of hosts and households is included, which allows for greater heterogeneity in the emulation of the city.

For the construction of the epidemiological model to be used in this paper, a deterministic model was used as a basis which allowed, among other things, the calculation of the basic reproductive number.

The ordinary differential equation model, which is used as a reference to create the numerical model used in this paper, is detailed below:

%The recruitment rate of new vectors makes the deterministic model a system of differential equations with stochastic forcing, due to the incorporation of climatic variables in modelling. When introducing an internal white noise, to represent demographic stochasticity, a Poisson approximation \cite{aparicio2001a} is used to solve the system of stochastic equations.
\begin{eqnarray}
\label{vsp}
 \frac{d {V_S}}{dt}  &=&  \Lambda (T,R,RH) - b \theta(T) p_V  \frac{{H_I}}{H}  {V_S}  - \mu_V(T) {V_S}  \\
\label{vep}
 \frac{d{V}_{E_1}} {dt} &=& b \theta(T) p_V  \frac{{V_S}}{H}  {H_I}  - \Bigg( k \sigma_V(T) + \mu_V(T) \Bigg) {{V}_{E_1}}  \\
\frac{d{V}_{E_2}}{dt} &=&  k \sigma_V(T){{V}_{E_1}}  -    \Bigg( k \sigma_V(T) + \mu_V(T) \Bigg) {{V}_{E_2}} \\
&\vdots & \nonumber \\
\frac{d{V}_{E_k}}{dt} &=& k \sigma_V(T){{V}_{E_{k - 1}}}   -    \Bigg( k \sigma_V(T) + \mu_V(T) \Bigg) {{V}_{E_k}} \\
\label{vip}
\frac{d {V}_I }{dt} &=& \sigma_V(T){{V}_{E_{k}}}  - \mu_V(T) {V_I} \\
\label{hsp}
\frac{d {H}_S  }{ dt } &=&  - b \theta(T) p_H {V_I} \frac{{H_S} }{H}  \\
\label{hep}
\frac{d {H}_E }{dt} &=&   b \theta(T) p_H {V_I} \frac{{H_S} }{H}  - \sigma_H  {H_E}  \\
\label{hip}
\frac{ d {H}_I  }{dt } &=&  \sigma_H  {H_E}   - \gamma {H_I}  + {\delta_I} \\
\label{hrp}
{H_R} &=&  H - {H_S} -  {H_I} 
\end{eqnarray}
Where $V_S$ and $V_I$ are the susceptible and infected vectors. 
The set of variables ($V_{E_1}, V_{E_2}, ..., V_{E_{k-1}},V_{E_k}$) 
correspond to successive exposed states of the vector, mimicking the time that needs the virus to disseminate from the midgut to the salivary glands inside the mosquito. The passage from one exposed state to another is done at a rate $k\sigma_V(T)$, being $k$ the shape parameter of the Gamma distribution function.
The rate $\sigma_V(T)$ depends on the mean temperature of the day.
The particular choice of a Gamma distribution is because it allows to mimic the effect of temperature by changing its parameter $k$, shortening or lengthening the exposed period accordingly. Additionally the gamma distribution, allows to pass from a differential-integral equation to a system of linear equations.
A susceptible vector becomes exposed at a $b \theta(T) p_V  \frac{{H_I}}{H}$ rate where  $b$, $\theta(T)$ and $p_V$ are the biting rate, the vector activity which depends on the average temperature and the probability of a susceptible vector becoming exposed when biting an infected host. 

As mentioned before, $V$ is the population of vectors. This population has a $\Lambda (T,R,RH)$ recruitment rate, which depends implicitly on the meteorological variables rainfall ($R$), average temperature ($T$) and relative humidity ($RH$). The value of $\Lambda (T,R,RH)$ is determined from the ecological model presented in this paper. Vector death occurs at a rate of $\mu_V(T)$ which is also a function of the average temperature.
The host population ($H$) is divided into susceptible ($H_S$), exposed ($H_E$), infected ($H_I$) and recovered ($H_R$) states. 

A fraction of susceptible hosts ($\frac{H_S}{H}$) become exposed at a rate $b \theta(T) p_H {V_I}$, where $p_H$ is the probability of a susceptible host becoming infected when is bitten by an infected vector. An exposed host becomes infected at a rate $\sigma_H$ and then recovered at a rate $\gamma$.

The term $\delta_I$ (Eq \ref{hip}) is the number of infected external hosts entering into the system at time t and is described in detail in \ref{serie_import}.

%Where $G_t(T) =  b \theta(T) p_V  \frac{{V_S}_t}{H}  {H_I}_t$. 
The basic reproductive number for the model described above can be derived from the model definitions (demonstration provided in \ref{derivationRzero}), therefore

\begin{equation}
R_0 = b^2 \ \theta(T) ^2 \ \frac{p_H p_V }{\mu_V(T) \ \gamma}  \left( \frac{k \ \sigma_V(T)  }{\mu_V(T) +  k \ \sigma_V(T) } \right)^k  \frac{V(t)}{H}
\end{equation}
%%%%%%%%%%%%%%%%%%%%%%%%%%%%%%%%%%%%%%%%
%%%%%%%%%%%%%%%%%%%%%%%%%%%%%%%%%%%%%%%%
%%%%%%%%%%%%%%%%%%%%%%%%%%%%%%%%%%%%%%%%
%%%%%%%%%%%%%%%%%%%%%%%%%%%%%%%%%%%%%%%%

\section{Ecological Model}
\label{ecological_model_app}
Starting from the \textit{Aedes aegypti} abundance model \cite{valdezLD2018}, an ecological model is developed to describe the population of vectors in the different epidemiological stages. In the present scheme (Fig.~\ref{ecological}), it is shown that the different stages of the vector depend on climatic variables, such as temperature (T), precipitation level (R) and relative humidity ($RH$). In particular, maturation stages of \textit{Aedes aegypti} are strongly related to climatic variations. Rainfall significantly affects the amount of eggs that hatch, while temperature and relative humidity mostly affect the maturation and mortality rates of the different stages of the vector.

To model the population dynamics of mosquitoes, we propose the following set of differential equations:
\begin{eqnarray}
\label{edp}
\frac{ dE_D }{dt} &=& \beta_{\mathrm{day}} \theta_0 (T) V - f(R){E_D} - \mu_{Dry} {E_D} \\
\label{ewp}
\frac{ d E_W }{ dt} &=& f(R){E_D}  - m_E(T,RH)C_G {E_W} - \mu_{Wet} {E_W} \\
\label{lp}
\frac{ dL }{dt}  &=& m_E(T,\mathrm{RH})C_G {E_W} - m_L(T,\mathrm{RH} )L \nonumber \\
 &-& \Big( \mu_L(T) + C_L(R,\mathrm{RH}) \Big) L\\
 \label{pp}
\frac{dP }{dt}  &=& m_L(T,\mathrm{RH}) L - m_P(T,\mathrm{RH}) P - \mu_P(T) P \\
\label{mp}
\frac{ dM }{ dt}  &=& m_P(T,\mathrm{RH}) P - m_M M  - \mu_M M \\
\label{vp}
\frac{dV}{dt}  &=& m_M M - \mu_V(T) V
\end{eqnarray}
Where ${E_D}$, ${E_W}$, $L$, $P$, $M$, $V$ are dried and wet eggs, larvae, pupae, young mosquitoes and adult female mosquitoes respectively.
In our models we consider that only half of the adult mosquitoes are female mosquitoes, a good approximation according to laboratory studies \cite{grech2010}. 
As a result of the ecological model, it can be determined that the rate of recruitment of new vectors (Eq.~\ref{vsp}) is equal to the rate of recruitment of adult female mosquitoes (Eq.~\ref{vp}), therefore
\begin{equation}
\label{rate_lambda}
    \Lambda (T,R,RH) = \frac{1}{2} m_M M
\end{equation}
So the adult female \textit{Aedes Aegypti} mosquitoes of the ecological model are the vectors of the epidemiological model.
\subsection{Dried and wet eggs}
The \textit{Aedes aegypti} life cycle starts when the mosquito lays its eggs on the waterline of (preferable artificial) water containers \cite{Harrington1998} at an oviposition rate given by the quantity $\beta_{\mathrm{day}} \theta_0 (T) V_t$. The parameter $\beta_{\mathrm{day}}$ is the oviposition rate of mosquitoes at optimal conditions (days$^{-1}$), while $\theta_0 (T)$ represent the effect of the temperature on this parameter. In the ecological model, all deposited eggs are considered as dry eggs (${E_D}$) and hatch when they flood due to the rain. It is proposed that if, on a given day, there is a precipitation of $R$ [mm], the fraction of ${E_D}$ that becomes wet eggs (${E_W}$) is given by Hill function:
\begin{equation}
\label{fraction_egg_app}
f(R) = 0.8 \frac{(R/R_{\mathrm{thres}})^5}{(1+R/R_{\mathrm{thres}})^5}
\end{equation}
where $R_{\mathrm{thres}}$ is the rainfall threshold and the factor $0.8$ is the maximum quantity of ${E_D}$ that may convert to ${E_W}$. 
The factor 0.8 appears due to a reduction in the hatching probability because of the presence of bacteria and larvae. This factor varies between 0.65 and 0.8 depending on the {\it Aedes} strain and local characteristics \cite{romeoaznar2013}. 
In this work a threshold value $R_{\mathrm {thres}} = 10.5$ mm is used, with no significant differences observed when other nearby values are used \cite{romeoaznar2013,valdezLD2018}. 

%%%%%%%%% DEFINIR NOTACION
%%%The egss mortality (dry and wet) occurs at a constant rate $\mu_E$.
The egss mortality occurs at a constant rate, $\mu_{Dry}$ and $\mu_{Wet}$, for dry and wet eggs respectively. 
Wet eggs (${E_W}$) hatch at a rate proportional to $m_E$ that depends on temperature. There is experimental evidence \cite{gillett_1955, gillett_1977} that suggests that the hatching process is delayed when the larval population increases. This process is known as the Gillett effect, and can be included in the model multiplying $m_E$ by:
\begin{equation}
\label{C_G_app}
C_G = \left\{
	       \begin{array}{ll}
		 1 - \frac{L}{K_L}      & \mathrm{if\ } L< K_L \\
		 0 & \mathrm{if\ }  L \geq K_L
	       \end{array}
	     \right.
\end{equation}
If the number the larvae ($L_t$) exceeds the larval capacity ($K_L$), then the wet eggs cannot hatch and ${E_W}$ cannot make a transition to the larval compartment. 

\subsection{Larvae}
\label{larvae_app}
The larvae mature at a rate $m_L$ that depends on temperature.
Larval population growth is restricted by intraspecies competition, which increases the larval mortality rate. This effect is included in the model by adding to the density-independent larval mortality rate $\mu_{L} (T)$, a term proportional to the larval population  \cite{romeoaznar2013} (see Section~\ref{mortality_aedes}). Then, larval mortality is modelled by:
\begin{eqnarray}
\label{C_L_app}
C_L = 1.5 \frac{L}{K_L}
\end{eqnarray}
Rainfall creates favourable breeding sites for larvae development, but evaporation tends to reduce these sites. Therefore, we propose that container's carrying capacity depends on the amount of water available $W (t)$, whose variation is defined as:

\begin{equation}
W(t+1) = \left\{
	       \begin{array}{ll}
		 0     & \mathrm{if\ } W(t) + \Delta(t) \leq 0 \\
		 W{\mathrm{max}} & \mathrm{if\ }  W(t) + \Delta(t) \geq W{\mathrm{max}}   \\
		 W(t) + \Delta(t) & \mathrm{otherwise } 
	       \end{array}
	     \right.
\end{equation}
Where  $\Delta(t) = R (t) - \mathrm{Evap}(t)$, being $R(t)$ the amount of rainfall on day $t$ and $\mathrm{Evap}(t)$ the daily evaporation for the same day. $W(t)$ can only increase to a maximum value $W_{\mathrm{max}}$, since at a certain water level, the containers or breeding sites overflow. In our model the evaporation rate $\mathrm{Evap}(t)$ is given by the expression\cite{romanenko1961,valipour2014}:

\begin{equation}
\mathrm{Evap}(t) = k_I (25{}^\circ \mathrm{C} + T(t) )^2(100 - \mathrm{RH(t)})
\end{equation}

\noindent where $\mathrm{RH(t)}$ and $T(t)$ (relative to  ${25}^{\circ}$C), are the average relative humidity and temperature at day $t$ and $k_I$ is the Ivanov constant.
We assume that the larval carrying capacity of each patch ($K_L^j$) depend on the number of houses ($n_h^j$) in that patch as:

\begin{equation}
\label{KL}
K_L (t) = K_{\mathrm{max}} \frac{W(t)}{W_{\mathrm{max}}} +1 
\end{equation}

\noindent where $K_{\mathrm{max}}$ is the maximum carrying capacity, $n_h^j$ is the number of houses in patch $j$-th, $W(t)$ is amount of water available, and can only increase to a maximum value $W_{\mathrm{max}}$. The term plus one is introduced to avoid divergences when $W(t) \rightarrow 0$.
In the case $W (t) \rightarrow 0$,  $K_L^j (t)$ will tend to the number of households, which implies that potentially a household can always contain a larvae. Here we consider $W_{\mathrm{max}} = 24$ mm and the value of $K_{\mathrm{max}}$, the maximum average number of mosquitoes per house per day, was chosen to be between 70 and 220; those values were fitted from experiments in reference \cite{valdezLD2018}. Those lasts values are also supported by the reported maximum numbers of {\it Aedes aegypti} larvae found in Oran city \cite{estallo2011}. 

According to our model, the larval capacity in each patch depends on the number of households, and the climatic conditions.

\subsection{Pupae, mosquitoes and vectors}
The mortality of pupae (${\mu}_P$) also depends on temperature \cite{romeoaznar2013}. 
After an average of $4$ days $(={m_M}^{-1})$ following pupation, the young mosquito is able to fly and bite.
Additionally, the mortality of young mosquitoes occurs at a constant rate ($\mu_M$).  

The recruitment rate of vectors ($\Lambda (T,R,RH)$) is given by the number of young mosquitoes ($M$) that mature and start to fly. The value of $\Lambda (T,R,RH)$ obtained with this ecological model is used as the input of the epidemiological model described above.
It is estimated that half of the mosquitoes are females\cite{grech2010}, the only ones that bite in order to produce eggs, and therefore the only ones able to transmit the virus.

Finally, we will assume that vector mortality occurs at a rate $\mu_V(T)$.
The dependence of the different rates (maturation and death) with temperature, is explained in the following Section.

Despite the evidence of vertical dengue transmission for \textit{Aedes aegypti} \cite{lima2018,espinosa2014,murillo2014}, in this work we considered it negligible. \\

\section{Climate effect in the \textit{Aedes aegypti}}
\label{clima_aedes}

\subsection{Temperature effect in the activity of \textit{Aedes aegypti} }
\label{aedes_activity}
In this work we propose that the activity of the mosquito is affected by the variation of the temperature. We define\cite{valdezLD2018} $a_0(T)$ function for the activity as:
\begin{eqnarray}
\label{activity_aed_T}
a_0(T) &=& 0.1137\Big(-5.4 + 1.8T - 0.2124T^2 \nonumber \\
&+& 0.01015T^3 - 0.0001515T^4\Big) 
\end{eqnarray}
Where $T$[${}^\circ \mathrm{C}$] is the average temperature. This function has a maximum for $T = 30$ $^o$ C .\\
\subsubsection{Oviposition}
\label{aedes_ovi}
In this work it is proposed that the number of dried eggs laid is equal to the oviposition rate $\beta_{\mathrm{day}}$ under optimal conditions of temperature. The factor $\theta_0(T)$, which allows to emulate the effect of temperature on oviposition is given by

\begin{equation}
\label{theta_0T}
\theta_0(T) = \left\{
	       \begin{array}{ll}
		 a_0(T)      & \mathrm{if\ } T \in (11.7, 37.2)[{}^\circ \mathrm{C}]  \\
		 0 & \mathrm{otherwise}
	       \end{array}
	     \right.
\end{equation}
Throughout its life, a female {\it Aedes aegypti} mosquito produces up to 60 eggs in optimal laboratory conditions \cite{sowilem2013}. Under this conditions the female  mosquito lives up to 30 days\cite{sowilem2013}. Therefore, in this paper we assume that the oviposition rate in optimal conditions has to satisfy the $\beta_{\mathrm{day}} > 2$ dry egg [day]$^{-1}$ condition.
\subsubsection{Biting rate}
\label{bite_rate}

For the transmission rate in vectors and hosts, and for the bite rate we propose to use the same normalized function $a_0$ that is used for oviposition. 

\begin{equation}
\label{thetaT}
\theta(T) = \left\{
	       \begin{array}{ll}
		 a_0(T)      & \mathrm{if\ } T \in (15, 35)[{}^\circ \mathrm{C}] \\
		 0 & \mathrm{otherwise}
	       \end{array}
	     \right.
\end{equation}

In this case we have a temperature threshold of 15$^\circ$ C so that there is a zero probability\cite{sowilem2013} of transmission.

\subsection{Effect of temperature in the latency period}
\label{latency_aedes}
From the experimental evidence it is proposed to use a gamma function. Its parameters are adjusted with an average rate that depends on temperature to build the survival function for the exposed state of the vector. Setting the shape parameter (k), different simulations were performed in order to qualitatively adjust the experimental data for \textit{Aedes aegypti} \cite{carrington2013}. 
\begin{equation}
\label{zeta}
\zeta(T) = \left\{
	       \begin{array}{ll}
		 e^{ -0.1659T + 6.7031 }      & \mathrm{if\ } k \leq 3 \\
		 e^{ -0.155T + 6.5031 } & \mathrm{if\ } k > 3
	       \end{array}
	     \right.
\end{equation}

\begin{figure}[ht]
\centering
\includegraphics[width = 8 cm]{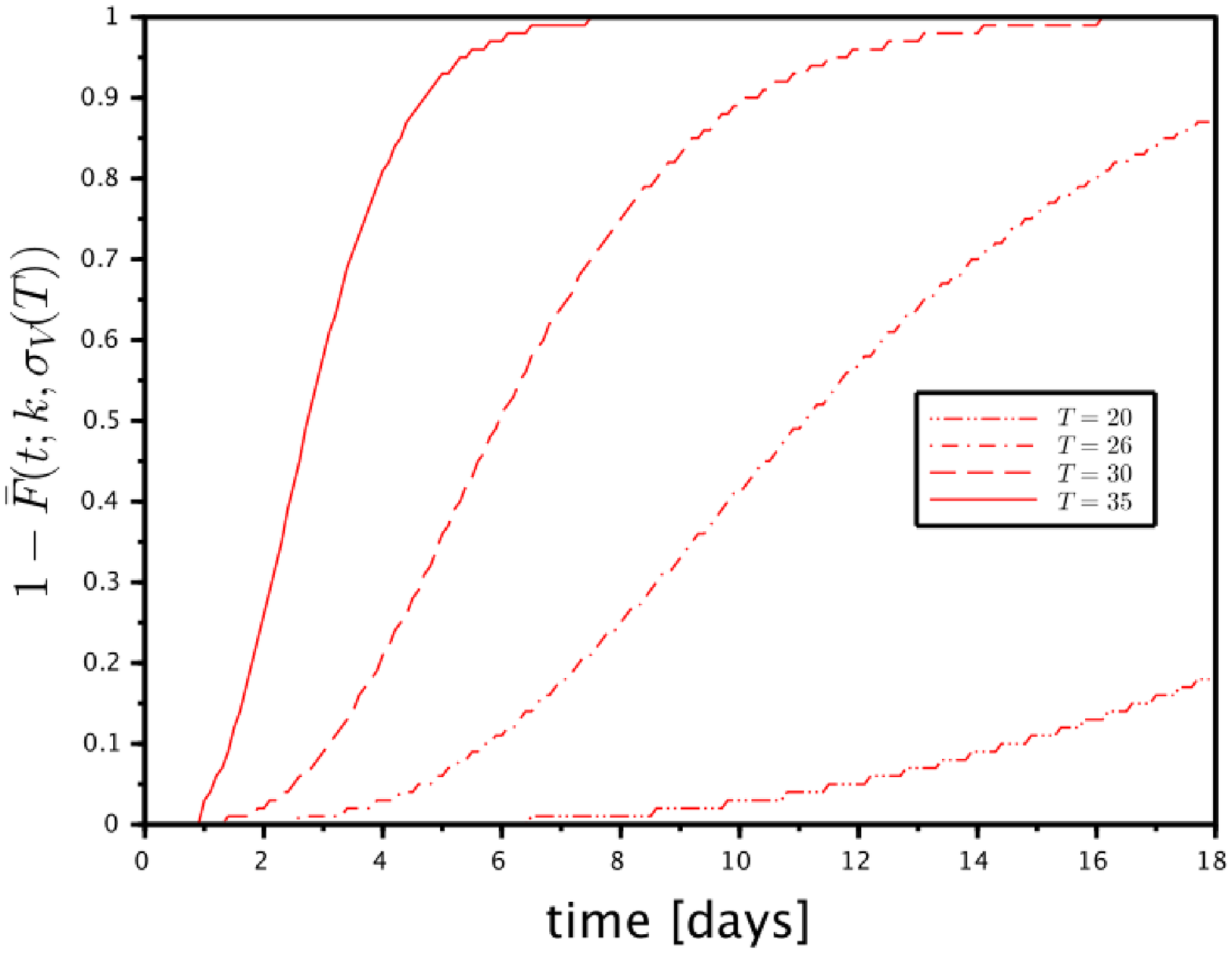}
\caption{Survival curves for the exposed state at different temperatures. It is appreciated that as the temperature drops, the fraction of mosquitoes that exceed the exposed state is lower. The different curves correspond to a shape parameter $k = 5$.  }
\label{survival_latency}
\end{figure}

\subsection{Maturation rates}
\label{maturation_aedes}
We define the maturation rates $m_X$ according to parameters listed in Table \ref{madurar}. Where $X= E, L$ and $P$ stand for egg, larvae and pupae. The maturation rate is defined\cite{solari2014} as:
\begin{equation}
\label{rate_maturate}
m_X = R_X \frac{T + T_0}{298}\frac{ exp\left[  \frac{\Delta H_A}{R_g} \left(  \frac{1}{298} - \frac{1}{T + T_0}\right) \right] }{1 + exp\left[  \frac{\Delta H_H}{R_g} \left(  \frac{1}{T_{1/2}} - \frac{1}{T + T_0}\right) \right]}
\end{equation}
\begin{table}[h!]
\centering
\caption{Parameters for the maturation rate}
\label{madurar}
\begin{tabular}{lllll}
\hline
X & $R_X$  & $\Delta H_A$ & $\Delta H_H$ & $T_{1/2}$ \\ \hline
E & 0.24   & 10978        & 100000       & 14184     \\
L & 0.2808 & 26018        & 55990        & 304.6     \\
P & 0.384  & 14931        & -472379      & 148       \\ \hline
\end{tabular}
\end{table}

Where $T$ is the mean temperature [($^\circ$C)], $T_0 = 273.15$ K is a factor to convert the temperature units from $^\circ$C to $K$, and $R_g$ is the universal gas constant. The set of parameters ($R_X$,$\Delta H_A$, $\Delta H_H$ and $T_{1/2}$) are shown in the table \ref{madurar} for $X= E, L$ and $P$. Moreover, we set the maturation rate coefficient for \textit{Aedes aegypti} larvae to be equal to zero for $T< 13.4$ $^o$C \cite{takashi2014}.

\subsection{Mortality rates}
\label{mortality_aedes}
We define\cite{sowilem2013} the mortality rates $\mu_{V}(T)$ as:
\begin{equation}
\mu_{V}(T) = \mu_{V_0} \eta(T)
\end{equation}
Here $\mu_{V_0}$ is the mortality of the \textit{Aedes aegypti} at a normal temperature of 22 $^o$C, the $\eta(T)$ [${}^\circ$C]  factor gives a variation for the mortality of the \textit{Aedes aegypti}. We define the prefactor:
\begin{eqnarray}
\eta(T) &=& 0.0360827 \Big(0.8692 - 0.159T + 0.01116T^2  \nonumber \\
										  &-&3.408\times 10^{-4} T^3 + 3.809\times 10^{-6}T^4\big) 
\end{eqnarray}
 For larvae and pupae the mortalities are defined\cite{solari2014} in the table \ref{mortality}.
 
\begin{table}[ht!]
\centering
\caption{Parameters for the mortality rate for L and P, where $T$ [${}^\circ$C] is the average daily temperature.}
\label{mortality}
\begin{tabular}{ll}
\hline
$\mu_L$ & $0.01 + 0.9725 e^{- \frac{T - 4.85}{2.70035}}$  \\ 
$\mu_P$ &  $0.01 + 0.9725 e^{- \frac{T - 4.85}{2.70035}}$  \\ \hline
\end{tabular}
\end{table}

For all mortality rates the average temperature is used.

%%%%%%%%%%%%%%%%%%%%%%%%%%%%%%%%%%%%%%%%
%%%%%%%%%%%%%%%%%%%%%%%%%%%%%%%%%%%%%%%%
%%%%%%%%%%%%%%%%%%%%%%%%%%%%%%%%%%%%%%%%
%%%%%%%%%%%%%%%%%%%%%%%%%%%%%%%%%%%%%%%%

\section{Climatic variables in the city of Oran}
\label{clima_oran}
\begin{figure}[ht!]
\centering
\includegraphics[width = 8 cm]{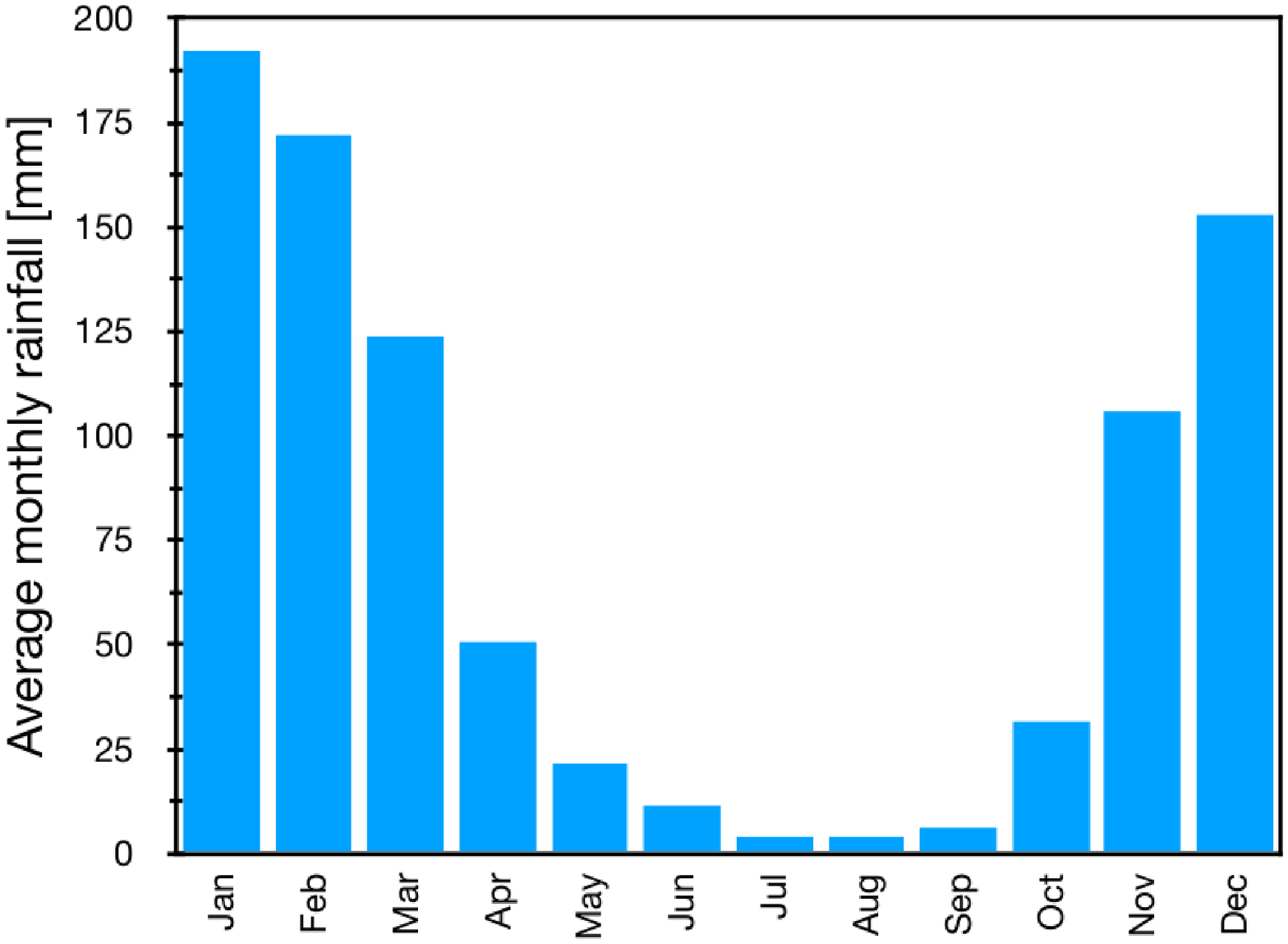}
\caption{Average monthly rainfall in the city of Oran for the period 2009-2017. }
\label{lluvia_oran}
\end{figure}

All the simulations were made with daily data of temperature, relative humidity and rainfall for the period 1998-2017, provided by the National Meteorological Service of Argentina \cite{smn}.
The city of San Ram{\'o}n de la Nueva Or{\'an} (city of Oran) is located in Northwest Argentina (with coordinates 23$^\circ$08'S 64$^\circ$20'W). Oran is a subtropical city with a seasonal climate with rains occurring mostly during the austral summer (see Fig.~\ref{lluvia_oran}).

\begin{figure}[ht!]
\centering
\includegraphics[width = 8 cm]{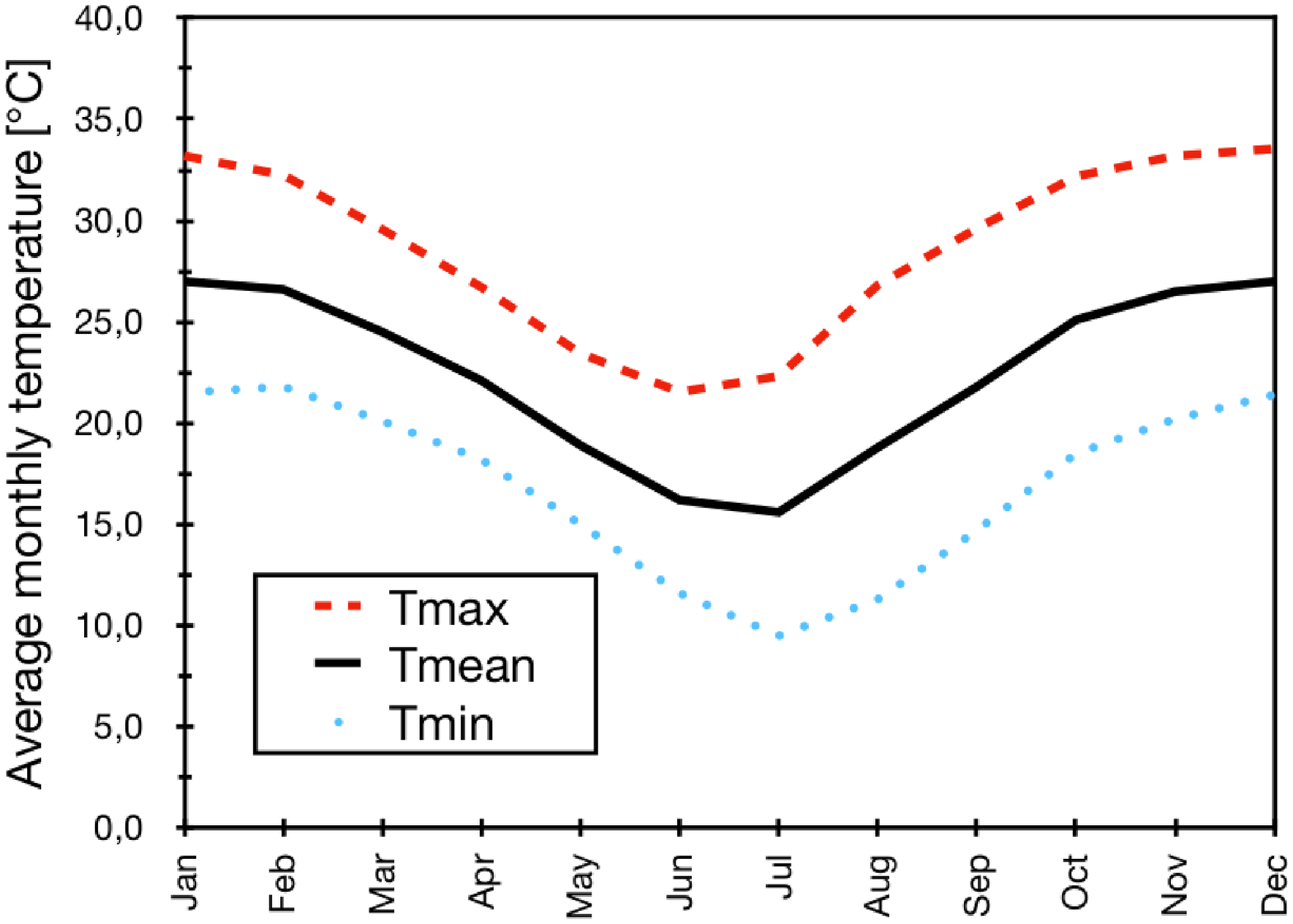}
\caption{Average of maximum, minimum and median temperatures in the period 2009-2017 for the city of Oran.}
\label{temp_oran}
\end{figure}

For the case of temperature, a seasonality weaker than rainfall's can be observed (see Fig.~\ref{temp_oran}).
For instance average temperature increases 60\% from July (minimum) to January (maximum) while for the case of rainfall the variability is greater than 4000\%.

% \begin{table}[]
% \centering
% \label{nuevo_cuadro}
% \caption{In the second column is the mean temperature for each season, in the third column is its corresponding standard error.}
% \begin{tabular}{cccc}
% \hline
% Season          & $T_{\mathrm{mean}}$ {[}${}^{\circ}$ C{]} & $\Delta T$ {[}${}^{\circ}$ C{]} & err \% \\ \hline
% Dec08 -- Apr09 & 26.01                                       & 2.96 & 11.36                              \\
% Dec09 -- Apr10 & 25.94                                       & 3.70 & 14.28                              \\
% Dec10 -- Apr11 & 25.05                                       & 3.90 & 15.56                              \\
% Dec11 -- Apr12 & 25.17                                       & 3.40 & 13.52                              \\
% Dec12 -- Apr13 & 25.60                                       & 3.45 & 13.46                               \\
% Dec13 -- Apr14 & 25.05                                       & 3.65 & 14.02                               \\
% Dec14 -- Apr15 & 25.19                                       & 2.71 & 10.83                              \\
% Dec15 -- Apr16 & 25.18                                       & 3.83 & 15.22                              \\ \hline
% \end{tabular}
% \end{table}

\section{Dengue in Bolivia}
\label{dengue_bolivia}
\begin{figure}[ht!]
\centering
\includegraphics[width = 9.5 cm]{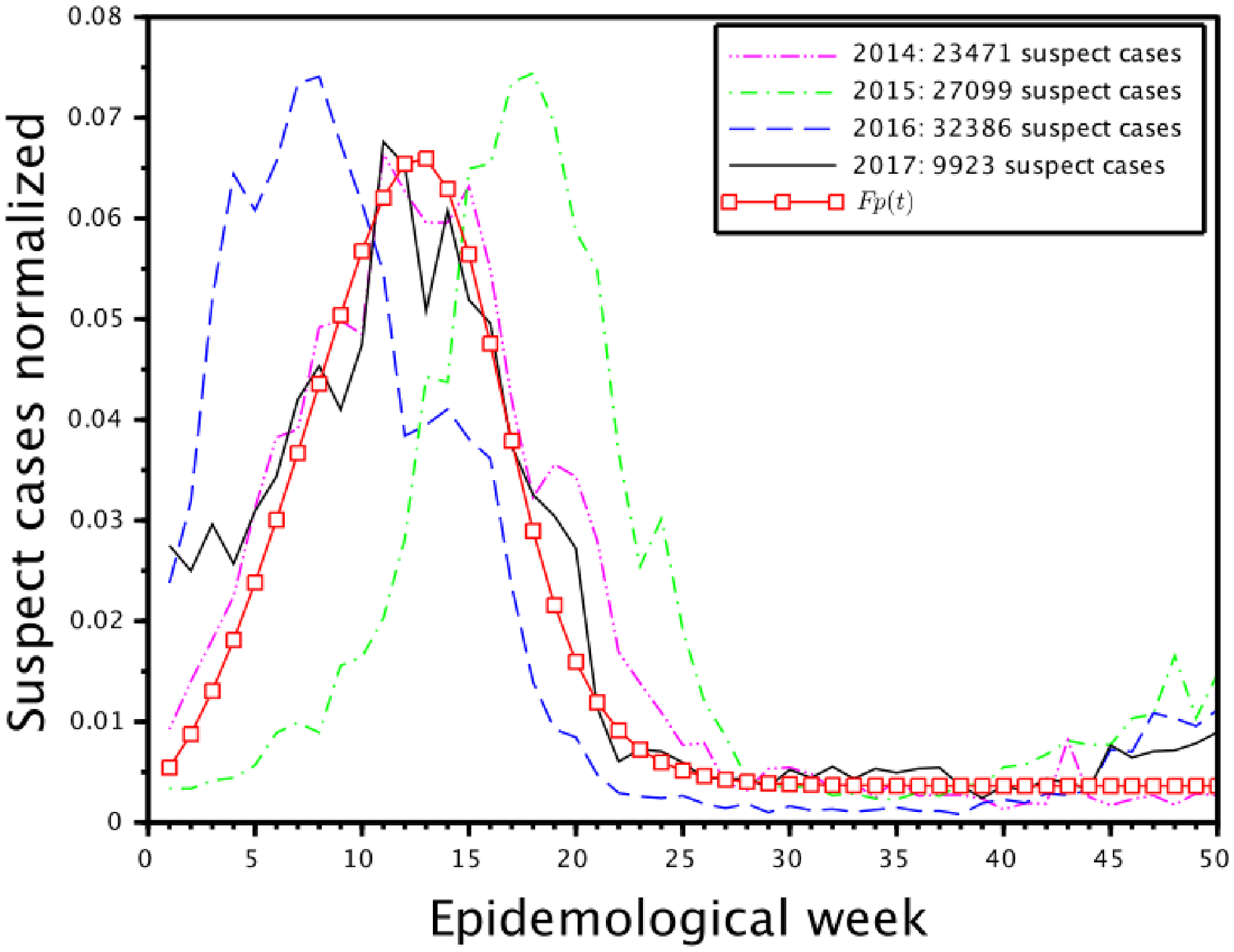}
\caption{Suspected cases of dengue in Bolivia, the values were normalised with the number of annual cases corresponding to each year.}
\label{reconstrucion}
\end{figure}

Suspected weekly cases of dengue in Bolivia are only available from 2014 at reference \cite{PAHO}. Therefore for the previous period 2010-2013, annual cases were distributed following the seasonality of those suspected cases. We propose the following function $Fp(t)$ that allows to reconstruct the series by epidemiological week:
\begin{equation}
Fp (t) = \alpha_i  \Bigg (2 + \frac{t^{1.5}}{e^{(t -15)/2  }+1} \Bigg) 
\end{equation}
This function $Fp(t)$ multiplied by the number of annual cases, allows to estimate the epidemic for the corresponding year, from January 2007 to  December 2013. The parameter $\alpha$ is adjusted so that the sum of all weekly cases is approximately equal to the corresponding annual cases. The calculated cases are rounded for each epidemiological week (Fig.~\ref{reconstrucion}).

\begin{figure}[ht!]
\centering
\includegraphics[width = 8 cm]{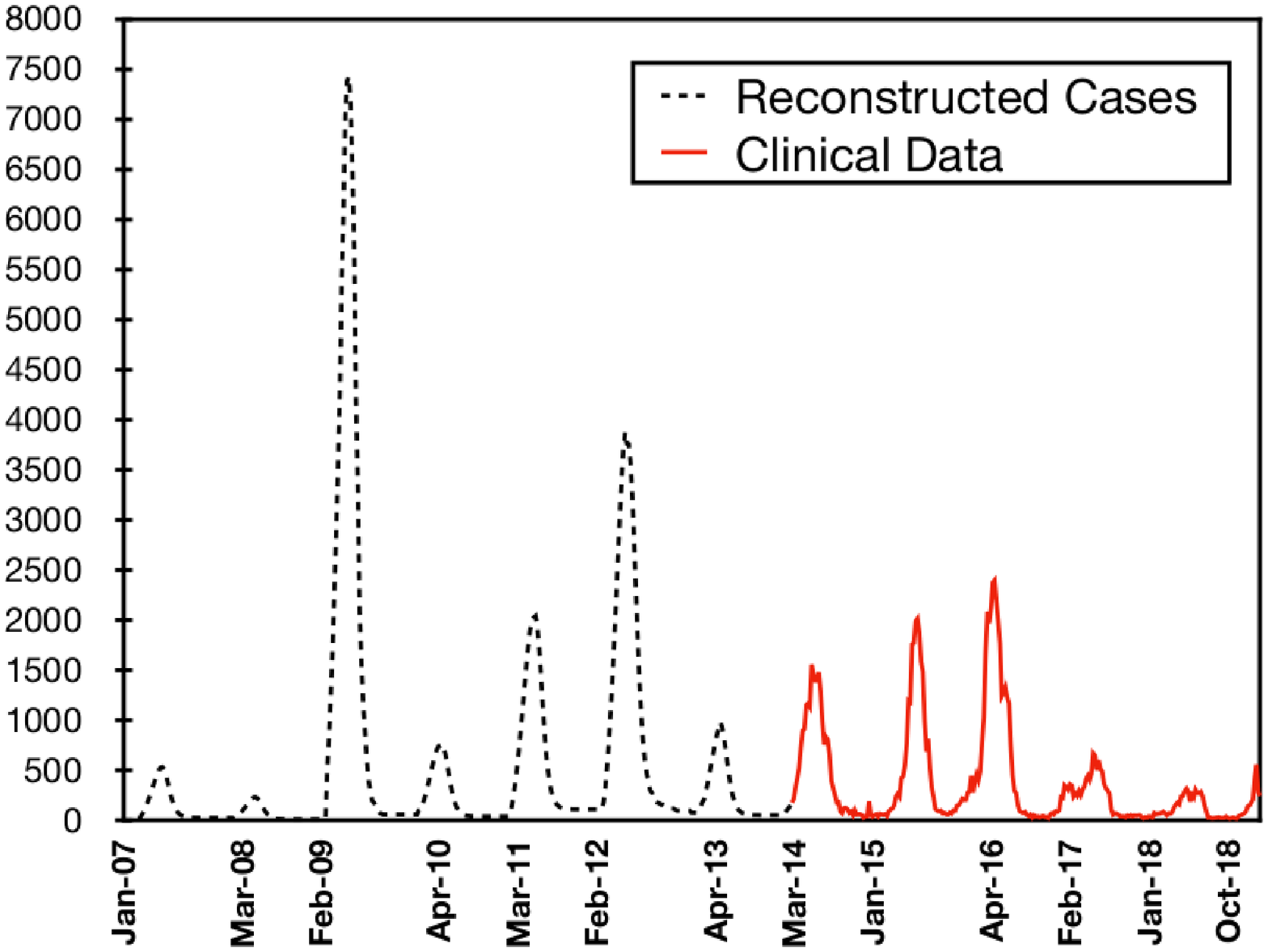}
\caption{The time series of dengue suspected cases in Bolivia from January 2007 to October 2018. The first period (black line) is re-constructed from the annual case data and the seasonality of dengue in Bolivia. For the second period (red line) the epidemiological weekly data series was obtained from the PLISA (Health Information Platform for the Americas) website. }
\label{imported}
\end{figure}

\section{Time series for the imported case model}
\label{serie_import}
\begin{figure}[ht!]
\centering
\includegraphics[width = 8.5 cm]{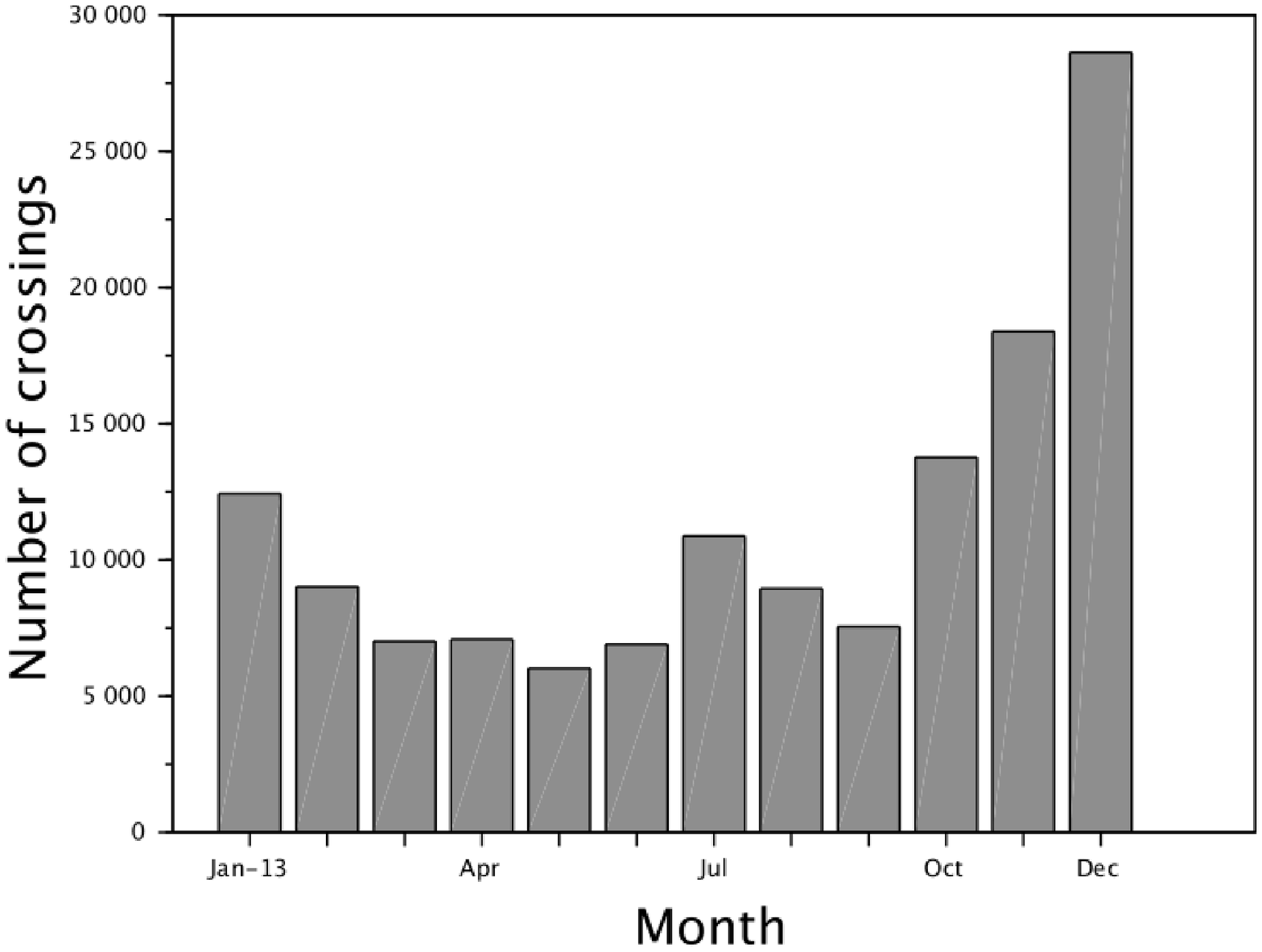}
\caption{Monthly people that entered Argentina during 2013 through the border crossing of Aguas Blancas. The data was taken from the website of the Gendarmer\'ia Nacional (Argentine National Gendarmerie).}
\label{cruces}
\end{figure}

We take as a reference the maximum number of people who entered Argentina through the Aguas Blancas border crossing in 2013 (see Fig.~\ref{cruces}). Accurate data on the number of people crossing the border is very difficult to obtain, due to structural issues and because there is a continuous and very high commercial and social exchange between countries.

\subsection{Imported cases flux proportional to the incidence in the source}
\label{importedflux}
We modelled the flux of imported cases using Eq.~(\ref{equation_imported}), where now $F_ {\mathrm {PS}}(t)$ is the number of people from Oran in contact with people from the endemic area of Bolivia, and $\frac{I_B(t)}{N_B}$ is the incidence of dengue observed in Bolivia.

We considered two cases for the flux of people from the non-endemic city visiting the endemic area ($F_ {\mathrm {PS}}(t)$). In the first case we assumed that this flux was constant along the year, while in the second case we assumed a flow concentrated in the rainy season, which coincides with the summer holidays. The constant flux was assessed as the average number of people crossing the Argentina-Bolivia border (Fig. \ref{cruces}). The concentrated flux was taken directly from Fig. \ref{cruces} which is in fact what is observed, i.e. higher flux during summer holidays. 

High and low intensity was set up by changing the value of the probability $p_i$ in Eq.~\ref{equation_imported}.

 \begin{figure}[ht]
\centering
\includegraphics[width = 8 cm]{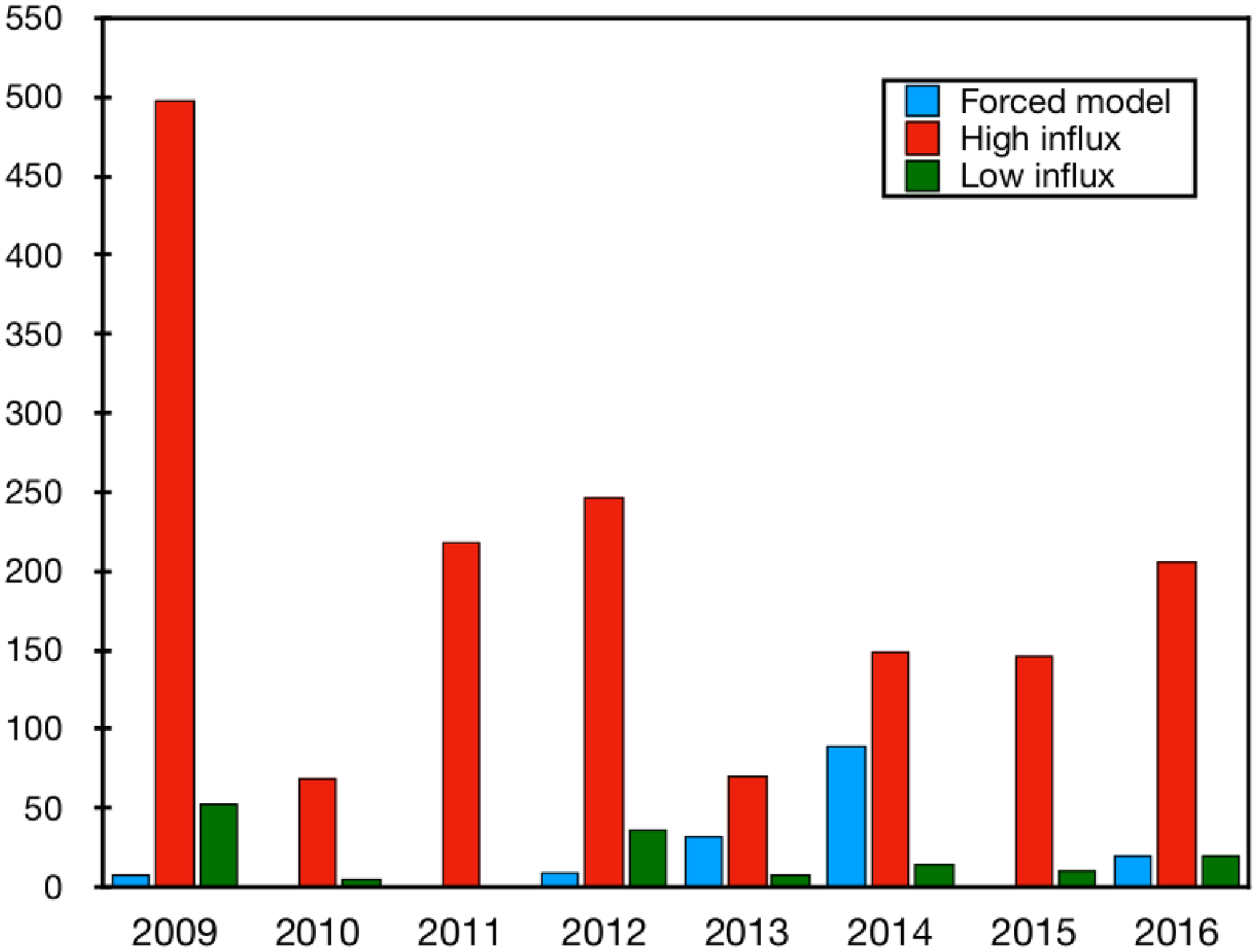}
\caption{Representative annual imported cases for the three scenarios: high constant influx (red bars), low constant influx (green bars), concentrated flux in the rainy season (sky-blue bars).}
\label{imported_cte_alto}
\end{figure}

\begin{figure}[ht!]
\centering
\begin{subfigure}
\centering
\includegraphics[width = 6.5 cm]{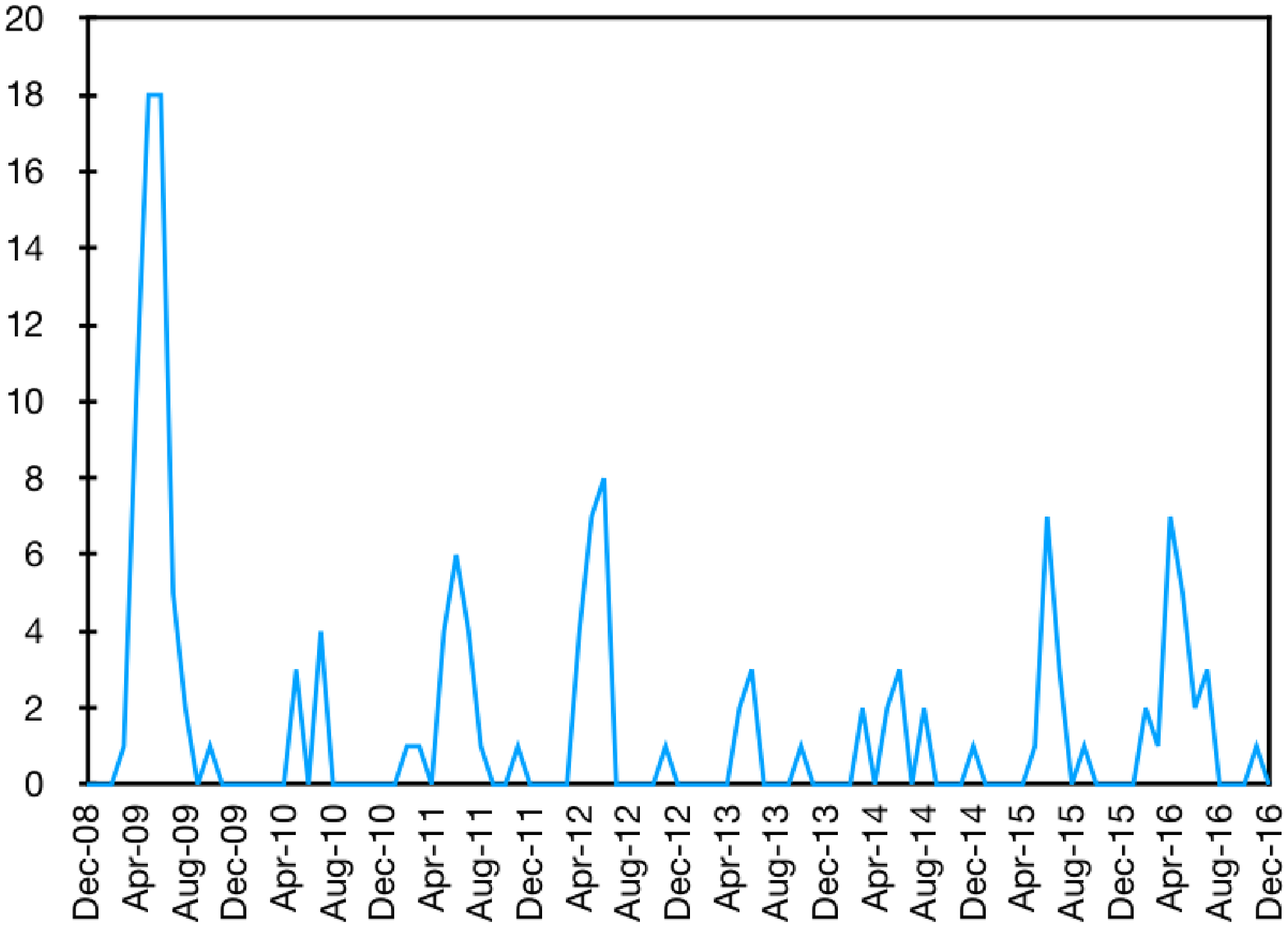}
\end{subfigure}
\begin{subfigure}
\centering
\includegraphics[width = 6.5 cm]{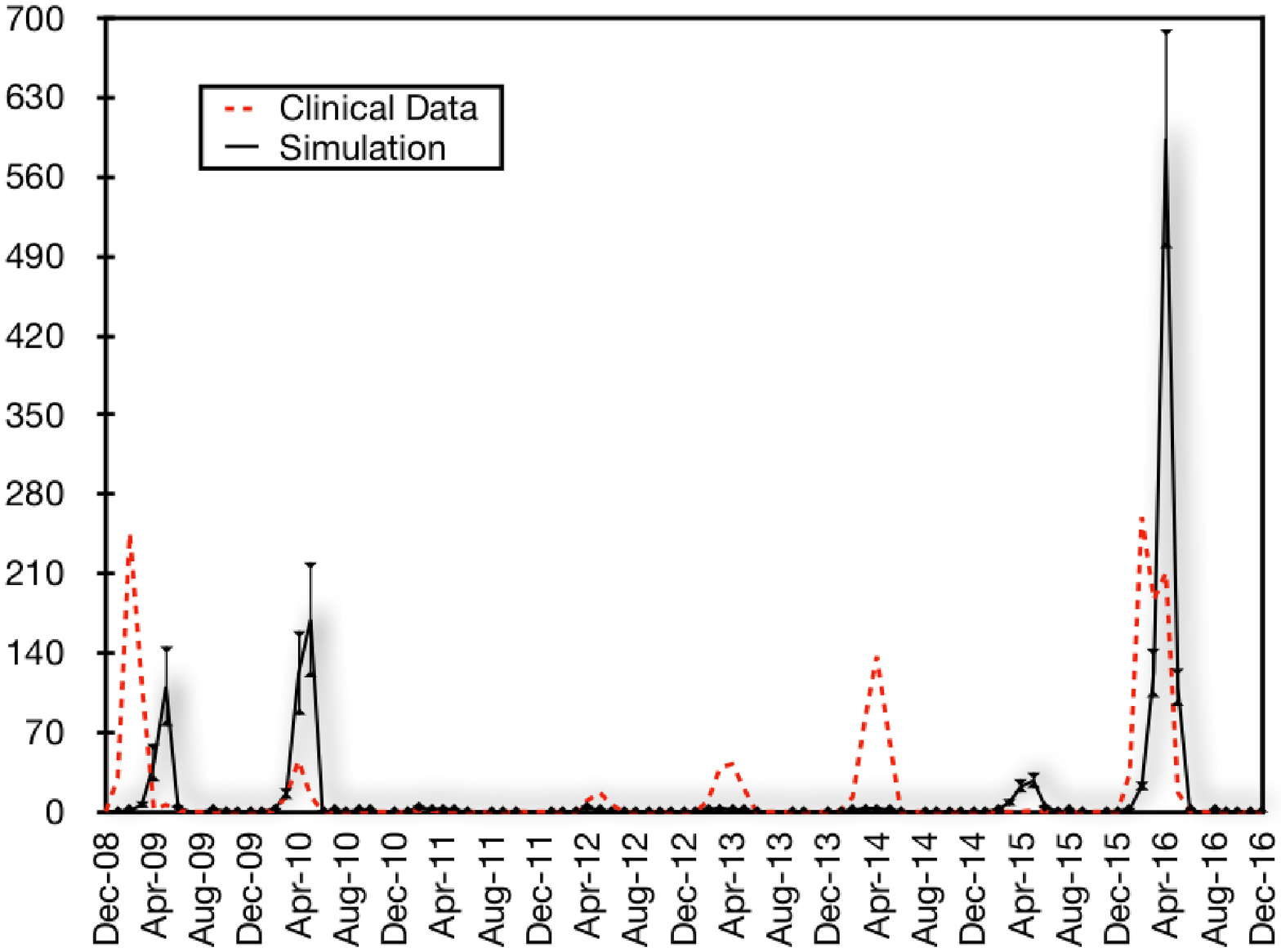}
\end{subfigure}
\caption{Left panel: representative time series of imported cases according to the model of low constant flux, which is used as input for the stochastic epidemiological model. The imported cases probability ($p_i$) was set as $0.25$. Right panel: comparison between observed and simulated dengue incidence using the model of low constant flux. The parameters of the \textbf{Table 1} are used to perform different simulations, they all start in January 2007 and end in December 2016. The series shown is an average of 100 realisations using different fluxes $\delta(t)$ in each case, the bars correspond to the standard error.}
\label{imported_constant}
\end{figure}

Simulations assuming a constant flux for $F_ {\mathrm {PS}}(t)$ in Eq.~\ref{equation_imported} are shown in Fig.~\ref{imported_constant} left panel. 
As imported time series cases was constructed assuming a constant flux of imported cases, the variability comes from the term proportional to the incidence in Bolivia.
They do not reproduce on average, the pattern of outbreaks observed in the data (Fig.~\ref{imported_constant} right panel), therefore assuming that imported cases are proportional to dengue incidence in Bolivia is not enough to reproduce the clinical data time series.

\begin{figure}[ht!]
\centering
\begin{subfigure}
\centering
\includegraphics[width = 6.5 cm]{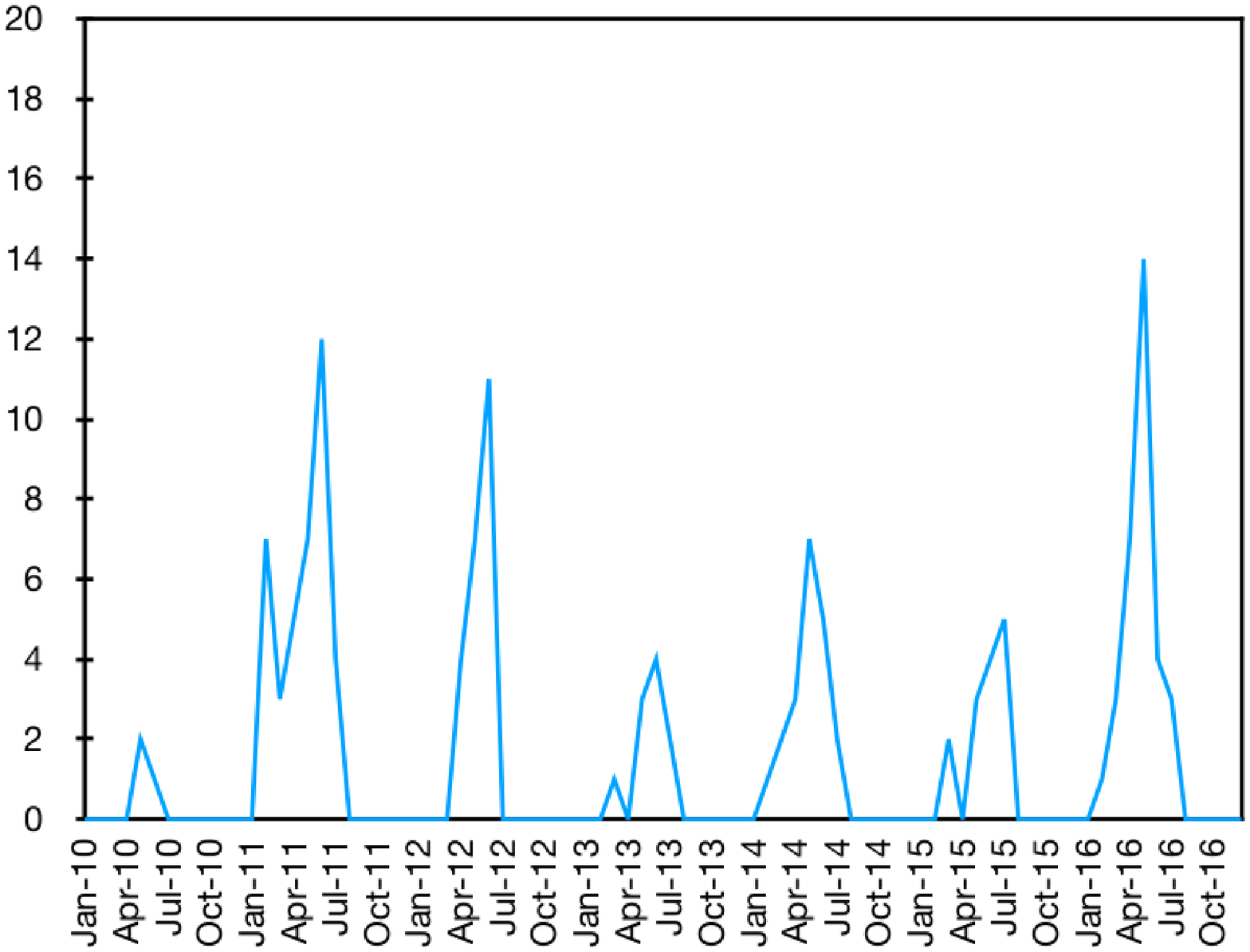}
\end{subfigure}
\begin{subfigure}
\centering
\includegraphics[width = 6.5 cm]{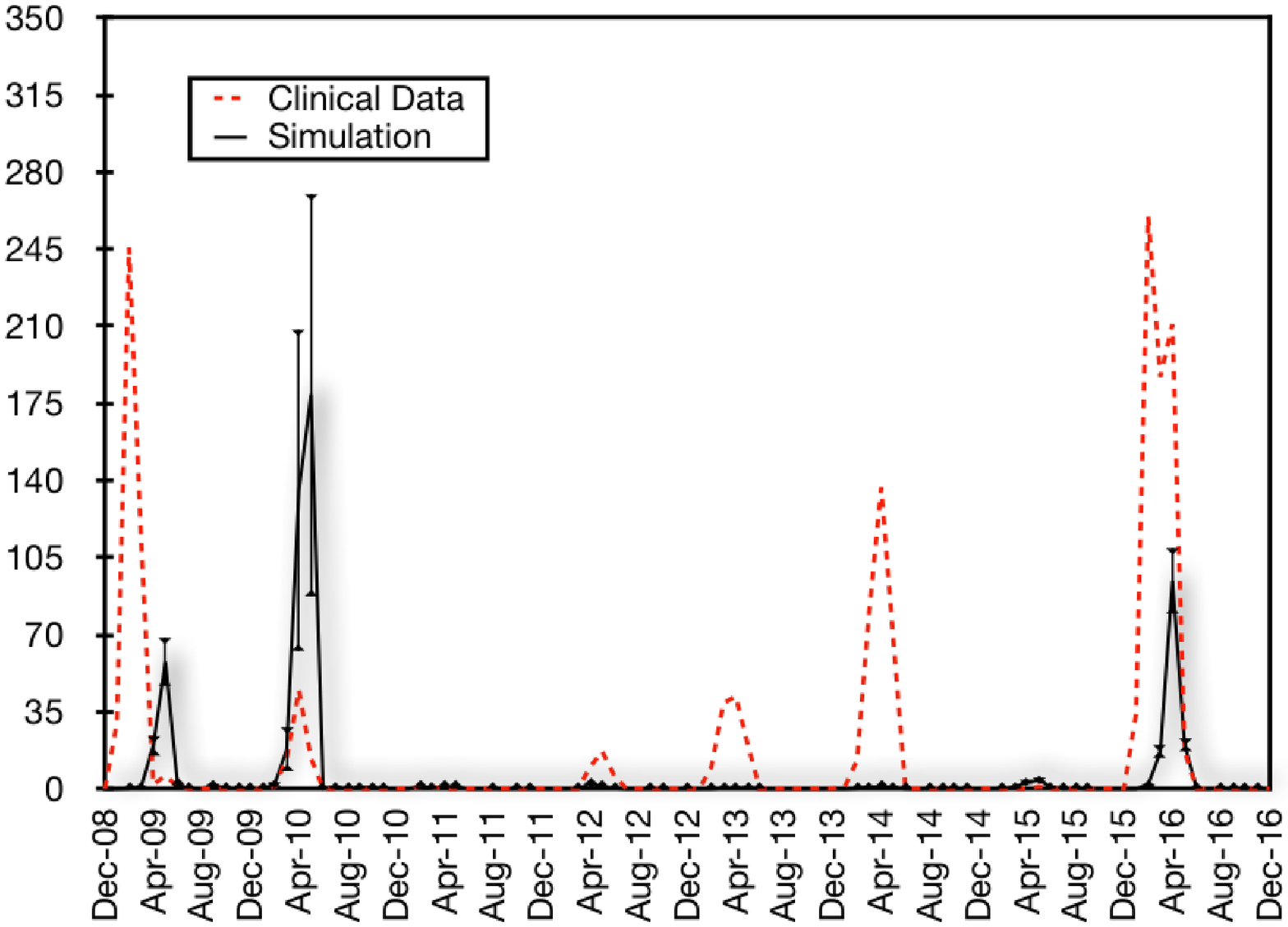}
\end{subfigure}
\caption{Left panel: Representative time series of imported seasonal cases from January to July for each year, used as input for the stochastic epidemiological model. The imported cases probability ($p_i$) was set in $0.25$. Right panel: Comparison between observed and simulated dengue incidence using the model of low seasonal flux. The parameters of the \textbf{Table 1} are used to perform different simulations, they all start in January 2007 and end in December 2016. The series shown is an average of 100 realisations using different fluxes $\delta(t)$ in each case, the bars correspond to the standard error. }
\label{imported_seas}
\end{figure}

Similarly using as input a seasonal influx of imported cases (Fig.~\ref{imported_seas} left panel), simulation averages do not reproduce either the pattern of outbreaks observed (Fig.~\ref{imported_seas} right panel). Therefore imported cases seasonality together with proportionality with dengue incidence in Bolivia, is not enough to reproduce clinical data variability. Other sources of variability are needed to reproduce the observed variability in dengue time series, as explained in the main text.

%%%%%%%%%%%%%%%%%%%%%%%%%%%%%%%%%%%%%%%%%%%%%%%%
%%%%%%%%%%%%%%%%%% VIEJO FLUJO CONSTANTE%%%%%%%%%%%%%%%%
%%%%%%%%%%%%%%%%%%%%%%%%%%%%%%%%%%%%%%%%%%%%%%%%
%% En todos los casos se respeta la cantidad de casos de bolivia, lo que cambia es que aqui p_i es constante todos los años y todos los meses
%%

%%%%%%%%%%%%%%%%%%%%%%%%%%%%%%%%%%%%%%%%%%%%%%%%
%%%%%%%%%%%%%%%%%%AGREGADO NUEVO%%%%%%%%%%%%%%%%
%%%%%%%%%%%%%%%%%%%%ESTACIONAL%%%%%%%%%%%%%%%%%%
%% En todos los casos se respeta la cantidad de casos de bolivia, lo que cambia es que aqui p_i tiene un comportamiento estacional es cero entre agosto y septiembre de todos los años, y para el resto de los meses (meses de transmisión) es distinto de cero
%%%%%%%%%%%%%%%%%%%%%%%%%%%%%%%%%%%%%%%%%%%%%%%%
%%%%%%%%%%%%%%%%%%%%%%%%%%%%%%%%%%%%%%%%%%%%%%%%
%%%%%%%%%%%%%%%%%%%%%%%%%%%%%%%%%%%%%%%%%%%%%%%%
%%%%%%%%%%%%%%%%%%%%%%%%%%%%%%%%%%%%%%%%%%%%%%%%
With this model for the imported cases the average of the simulations qualitatively reproduces the observed peaks in 2010 and 2016 but not the small outbreaks in between, neither using values of $p_i$ close to one.

If the flux of people visiting  the endemic region is distributed only during the rainy season of the non-endemic area, no qualitatively differences were observed. 

\section{Simulations}
\label{simulations}
In our model, density dependent processes take place during finite periods of time and involve many different events such as birth, death, change of states, among others. In some models it is possible to simplify the calculation of the number of cases for each event \cite{javier2015}, however in this work this is not possible because there are interlinked events.
Starting with the set of Eqs.~(\ref{vsp} - \ref{hrp}) and (\ref{edp} - \ref{vp}), we proceed to identify the number of cases, $X_i(t)$, corresponding to each event, $i$, in the model for the next time step $t + \Delta t$. We consider them as random variables defined through a (approximate) Poisson distribution. 
Then the realizations for each event are obtained as
\begin{equation}
\label{poisson}
    X_i(t + \Delta t) = \mathrm{Poisson} ( \lambda_i \Delta t )
\end{equation}
\noindent where $\mathrm{Poisson} ( \lambda_i \Delta t )$ is a (positive integer) number sampled from a Poisson distribution with parameter $\lambda_i$. For $ \lambda_i > 5$ the Poisson distribution is approximately a normal distribution with mean and variance equal to $\lambda_i$. In this case the Eq.~\ref{poisson} could be approximated with
\begin{equation}
    dX_i(t + d t) =  \lambda_i(X_i,t) dt + \sigma_{0,1}\lambda_i(X_i,t) dt
\end{equation}
where $\sigma_{0,1}$ is a random variable with normal distribution with zero mean and variance equal to one.

For the events considered in our epidemiological and ecological models (see Table \ref{states}), we calculate the corresponding number of occurrences, at each $\Delta t =1 $ day, using the Eq.~(\ref{poisson}). Certainly we must observe the balance among different populations, as shown in the corresponding diagrams (see Fig.~\ref{modelo_epi} and \ref{ecological}), updating the value of each variable at every time step.

%%% A particular case results from calculating the amount of dead vectors. In order to optimize and keep the balance of events, first the number of dead vectors is calculated in time $t + \Delta t$ , then using a multinomial function it is distributed among the different states of the vector ($V_S, V_{E_1}, ..., V_{E_k},V_I$).

\begin{table}[]
\centering
\caption{Different events that appear in the ecological and epidemiological model}
\begin{tabular}{llc}
\hline
Event                                      & $X_i$                   & $\lambda_i$                                    \\ \hline
Recruitment of a vector                    & $X_{\Lambda}$           & $m_M(M)$                                       \\
Death of a vector                          & $X_{\mu_V}$             & $\mu_V(T) V$                                   \\
New exposed vector                         & $X_{V_{E_1}}$           & $b \theta(T) p_V  \frac{{H_I}}{H}  {V_S}$      \\
i-th new exposed vector (i\textgreater{}2) & $X_{V_{E_i}}$           & $k\sigma_V (T) V_{E_{i-1}}$                    \\
New infected vector                        & $X_{V_I}$               & $k\sigma_V (T) V_{E_k}$                        \\
New exposed host                           & $X_{H_E}$               & $b \theta(T) p_H {V_I} \frac{{H_S} }{H}$       \\
New infected host                          & $X_{H_I}$               & $ \sigma_H  {H_E}$                             \\
New imported case                          & $X_{\delta_I}$          & $p_i F_ {\mathrm {PS}} (t) \frac{I_B(t)}{N_B}$ \\
New recovered host                         & $X_{H_R}$               & $\gamma H_I$                                   \\
Oviposition                                & $X_{Ov}$                & $\beta_{\mathrm{day}} \theta_O (T) V$          \\
Death of a dried egg                       & $X_{ \mu_{ E_{Dry} } }$ & $ \mu_{ E_{Dry} } E_{D} $                      \\
Death of a wet egg                         & $X_{ \mu_{ E_{Wet} } }$ & $ \mu_{ E_{Wet} } E_{W} $                      \\
Maturation of a wet egg                    & $X_{m_{E}}$             & $m_E(T, \mathrm{RH}) C_G {E_W}$                \\
Maturation of a larva                      & $X_{m_{L}}$             & $m_L(T,\mathrm{RH}) {L}$                       \\
Death of a larva                           & $X_{\mu_L}$             & $\Big( \mu_L(T) + C_L(R,\mathrm{RH}) \Big) L$  \\
Maturation of a pupa                       & $X_{m_{P}}$             & $m_P (T,\mathrm{RH}) {P}$                      \\
Death of a pupa                            & $X_{\mu_P}$             & $m_P(T,\mathrm{RH}) {P}$                       \\
Death of young mosquito                    & $X_{\mu_P}$             & $\mu_{ M } M$                                  \\ \hline
\end{tabular}
\label{states}
\end{table}
%%The results obtained were grouped by epidemiological week or by month for convenience to show results and to compare with epidemiological data.\\

The libraries used for random number generation are those of the Numerical Recipe. All programs were developed in C language and are available, along with the data series (meteorological, social information, clinical data) used, in the following link: {\url{ https://github.com/javoxa/ORAN\_PNTD}}.

\section{Deterministic epidemiological model}
\label{det_epidemiological_model}
We divided a constant host human population ($H$), in four different sub-populations according to its disease state, i. e. $H_S$, $H_E$, $H_I$ and $H_R$, the  susceptible, exposed, infected and recovered hosts populations respectively. For mosquitoes we used an ecological model for the \textit{Aedes} population ($V$), which depends on different climatic variables. In the same way we divided the vector population in $V_S$, $V_E$ and $V_I$ which are the susceptible, exposed and infected vector populations. The epidemiological model can be expressed as the following system of differential equations:
\begin{eqnarray}
\label{vs}
\dot{V}_S (t)&=& \Lambda (t;T,R,Hum) - G(t;T) - \mu_V(T) V_S(t) \\
\label{ve}
\dot{V}_E (t)  &=&   G(t;T)  - EV(t;V_E,T) - \mu_V(T) V_E(t) \\
\label{vi}
\dot{V}_I (t) &=& EV(t;V_E,T)  - \mu_V(T) V_I(t) \\
\label{hs}
\dot{H}_S (t) &=&  - \beta_r \theta(T) b_H V_I (t) \frac{H_S (t) }{H} \\
\label{he}
\dot{H}_E (t) &=&   \beta_r \theta(T) b_H V_I (t) \frac{H_S (t) }{H} - \sigma_H  H_E(t) \\
\label{hi}
\dot{H}_I (t) &=&  \sigma_H  H_E(t)  - \gamma H_I(t) + \delta_I(t) \\
\label{hr}
H_R (t) &=&  H - H_S(t) -  H_I(t) 
\end{eqnarray}
Where $G(t;T)$ is the rate at which vectors acquire the disease after biting an infected host, and is given by:
\begin{equation}
G(t;T) =  \beta_r \theta(T) b_V  \frac{V_S (t)}{H}  H_I (t)  \nonumber\\
\end{equation}
where the parameters $b_H$, $\sigma_H^{-1}$ and $\gamma^{-1}$ are the probability of contagious, the latent and infectious periods of the infected host respectively. In the vector evolution equations, the parameter $b_V$ is the probability of contagious, $\beta_r$ is the biting rate per day in optimum conditions, and $\theta(T)$ the effect of temperature on the mosquito bite rate. All transitions occur at a constant rate except for $V_{E \rightarrow I}$ transition. $EV(t;V_E,T)$ is the amount of exposed vectors that become infectious per unit time. This function can be obtained from a Volterra formulation of the SEI model, where the latent period distribution appears indirectly: 
\begin{eqnarray}
EV(t;V_E,T) = \lambda(t;T) V_E(t) \nonumber \\
+ \int_0^t \lambda(t;T) G(t;T) \bar{F} (t-s;T) e^{-\mu_V(T) (t-s)}ds \nonumber \\
 - \lambda(t;T)\int_0^t  G(t;T) \bar{F} (t-s; T) e^{-\mu_V(T) (t-s)}ds \nonumber
\end{eqnarray}
The function $\bar{F}(t;T)$ is the complementary cumulative distribution, also known as the survival function, and gives the probability that an individual exposed at $t = 0$ remains latent at time $s$. The fraction $\bar{F} (t-s;T)$ represent the latent infecteds at time $s$ that survive until time $t$. When the survival function is an exponential of the form $e^{-\lambda(t;T)  t}$, then $EV(t;V_E,T) = \lambda(t;T) V_E(t)$ ; with  $\lambda(t;T)^{-1}$  the mean latency period of the exponential distribution for vectors.

In this research work a gamma distribution function is considered for the survival function $ \bar{F} (t;T)$, fitting its parameters according to the latency survival curves for \textit{Aedes aegypti}. For the gamma function $\Gamma (t; k, \lambda(t;T ))$ with $\lambda(t;T) = \sigma_V(T)/k$ and $\sigma_V(T)$ the mean latency period, an integer $k$ is chosen so that Eq.~(\ref{ve}) can be rewritten as a set of differential equations:
\begin{eqnarray*}
\dot{V}_{E_1}(t) &=& G(t;T) - \Bigg( k \sigma_V(T) + \mu_V(T) \Bigg) {V}_{E_1} (t) \\
\dot{V}_{E_2} (t) &=&  k \sigma_V(T){V}_{E_1} (t)   -    \Bigg( k \sigma_V(T) + \mu_V(T) \Bigg) {V}_{E_2} (t) \\
&\vdots & \nonumber \\
\dot{V}_{E_k} (t) &=& k \sigma_V(T){V}_{E_{k - 1}} (t)   -    \Bigg( k \sigma_V(T) + \mu_V(T) \Bigg) {V}_{E_k} (t) 
\end{eqnarray*}
And satisfying the condition $V_{E} (t) = {\displaystyle \sum_{i=1}^{ k} V_{E_i}(t)} $. If $EV(t;V_E,T)$ is replaced by this set of differential equations, then it becomes necessary to replace the Eq.~(\ref{vi}) by:
\begin{equation}
 \dot{V}_I (t)  = k \sigma_V(T)  {V}_{E_k}  - \mu_V(T) V_I(t) 
\end{equation}
Finally $\Lambda (t;T,R,Hum)$ is the recruitment rate of new vectors. Given that vertical transmission for vectors it is not considered, that rate only depends on temperature ($T$), rain ($R$) and humidity ($Hum$). That function together with the infected hosts entering the system $\delta_I(t)$, also referred to as imported cases, will be described in the following sections "Ecological model" and "Imported cases model" respectively.

The basic reproductive number of our epidemiological model can be written as:
\begin{equation}
\label{R_0_app}
R_0 = \beta_r^2 \theta(T) ^2 \frac{b_h b_v }{\mu_V \gamma}  \left( \frac{k \sigma_V  }{\mu_V(T) +  k\sigma_V } \right)^k  \frac{V(t)}{H}
\end{equation}
where $\left( \frac{k \sigma_V (T)}{\mu_V(T) + k \sigma_V } \right)^k$ is the fraction of mosquitoes that survive the different $k$ latency stages. Not considering vital dynamics for hosts, the fraction of exposed hosts that survive latency is equal to 1. In our model we will always use a fixed value ($k =$ constant) for the different latent stages of the mosquito in all simulations. In this way, $R_0$ is proportional to the number of mosquitoes per human. When the mosquito population is very low, the probability of epidemic is also low.
\section{Mathematical derivation}
\subsection{$EV(t;V_E,T)$}
We wrote the Volterra Integral equation for exposed vectors:
\begin{equation}
\label{V_E}
V_E(t) = V_E(0) \bar{F}(t;T) e^{-\mu_V(T) t} + \int_0^t {  G(t;T) U(t-s;T) ds }
\end{equation}
where $G(t;T)  = \beta_r \theta(T) b_V  \frac{V_S (t)}{H}  H_I (t) $ and  $U(t-s;T) = \bar{F} (t-s;T) e^{-\mu_V(T) (t-s)}$.
The fraction $\bar{F} (t-s;T)$ represent the latent infecteds at time $s$ that survive until time $t$. \\
We derive the Eq.~(\ref{V_E}) to obtain a differential equation, then:
\begin{eqnarray}
\label{der_1}
\frac{d V_E(t)}{dt} = V_E(0) \frac{d\bar{F}(t;T) }{dt}e^{-\mu_V(T) t}  - \mu_V(T) V_E(0) \bar{F}(t;T) \nonumber \\ 
e^{-\mu_V(T) t} + G(t;T) U(0;T)
+ \int_0^t {  G(t;T) \frac{d U(t-s;T)}{dt} ds } \nonumber
\end{eqnarray}
Where
\begin{eqnarray*}
\frac{d U(t-s;T)}{dt}  = \frac{d\bar{F}(t-s;T)}{dt} e^ {-\mu_V(T) (t-s)} -  \mu_V(T) \\ 
\bar{F}(t-s;T) e^{-\mu_V(T) (t-s)}
\end{eqnarray*}
The failure rate for latency survival is:
\begin{equation}
\lambda(t;T) = \frac{pdf(t;T)}{\bar{F}(t;T)}
\end{equation}
Where $pdf(t;T)$ is the distribution density of the mean latent period. We can rewrite the failure rate so that it is defined only by the supervening function, then:
\begin{equation}
\frac{d\bar{F}(t-s;T)}{dt} = - \lambda(t;T) \bar{F}(t;T)
\end{equation}
So we can rewrite $\frac{d U(t-s;T)}{dt} $ in terms of the failure rate:
\begin{eqnarray*}
\frac{d U(t-s;T)}{dt}  = - \Big( \lambda(t;T)  +\mu_V(T) \Big) \bar{F}(t-s;T) e^{-\mu_V(T) (t-s)} \\ 
= - \Big( \lambda(t;T)  +\mu_V(T) \Big) U(t-s;T)
\end{eqnarray*}

Rewriting the Eq.~(\ref{der_1}) we obtain:
\begin{eqnarray*}
\frac{d V_E(t)}{dt} &=& - \lambda(t;T) V_E(0) \bar{F}(t;T) e^{-\mu_V(T) t}  \\
&-& \mu_V(T) V_E(0) \bar{F}(t;T) e^{-\mu_V(T) t} \\ 
&+& G(t;T) U(0;T) - \int_0^t {\lambda(t;T) G(t;T) U(t-s;T) ds }\\
&-& \int_0^t {\mu_V(T)  G(t;T) U(t-s;T) ds } 
\end{eqnarray*}
As $\mu_V(T)$ is implicitly time-dependent and, since $T= T(t)$, it cannot be operated like $\lambda(t;T)$, therefore:

\begin{eqnarray*}
\frac{d V_E(t)}{dt} &=& G(t;T) U(0;T)  - \lambda(t;T) V_E(0) \bar{F}(t;T) e^{-\mu_V(T) t} \\
&-& \mu_V(T) V_E(0) \bar{F}(t;T) e^{-\mu_V(T) t} \\
&-&\int_0^t {   \lambda(t;T) G(t;T) U(t-s;T) ds }\\
&-& \int_0^t {  \mu_V(T)  G(t;T) U(t-s;T) ds }\\
&-& \lambda(t;T)  \int_0^t {   G(t;T) U(t-s;T) ds }\\ &-&\mu_V(T) \int_0^t {   G(t;T) U(t-s;T) ds } \\
&+& \lambda(t;T)  \int_0^t {   G(t;T) U(t-s;T) ds }\\
&+& \mu_V(T) \int_0^t {   G(t;T) U(t-s;T) ds } 
\end{eqnarray*}
Re-ordering
\begin{eqnarray*}
\frac{d V_E(t)}{dt} &=& G(t;T) U(0;T)  - \lambda(t;T)  \Bigg( V_E(0) \bar{F}(t;T) e^{-\mu_V(T) t} \\
&+& \int_0^t {   G(t;T) U(t-s;T) ds } \Bigg) \nonumber \\
&-&  \mu_V(T) V_E(0) \Bigg(  V_E(0) \bar{F}(t;T) e^{-\mu_V(T) t}  \\
&+& \mu_V(T) \int_0^t {   G(t;T) U(t-s;T) ds } \Bigg) \nonumber \\ 
&+& \lambda(t;T)  \int_0^t {   G(t;T) U(t-s;T) ds } \\
&-&  \int_0^t {   \lambda(t;T) G(t;T) U(t-s;T) ds }   \nonumber \\
&+&  \mu_V(T) \int_0^t {  G(t;T) U(t-s;T) ds } \\
&-&  \int_0^t {  \mu_V(T)  G(t;T) U(t-s;T) ds } 
\end{eqnarray*}
Introducing (\ref{V_E}) in the last equation we get: 
\begin{eqnarray*}
\frac{d V_E(t)}{dt} &=& G(t;T) U(0;T)   -  \mu_V(T) V_E(t) \nonumber \\ 
&-&  \lambda(t;T)  V_E(t) + \lambda(t;T)  \int_0^t {   G(t;T) U(t-s;T) ds } \\
&-&  \int_0^t {   \lambda(t;T) G(t;T) U(t-s;T) ds }   \nonumber \\
&+&  \mu_V(T) \int_0^t {  G(t;T) U(t-s;T) ds } \\
&-&  \int_0^t {  \mu_V(T)  G(t;T) U(t-s;T) ds } 
\end{eqnarray*}
Where we define:
\begin{eqnarray*}
EV(t;T) &=& \lambda(t;T)  V_E(t) -  \lambda(t;T)  \int_0^t {   G(t;T) U(t-s;T) ds } \nonumber \\ 
&+&  \int_0^t {   \lambda(t;T) G(t;T) U(t-s;T) ds }   \nonumber \\
&-&  \mu_V(T) \int_0^t {  G(t;T) U(t-s;T) ds } \nonumber \\ 
&+& \int_0^t {  \mu_V(T)  G(t;T) U(t-s;T) ds } 
\end{eqnarray*}
Since no age structure is used for the vectors, $\mu_V(T) ^{-1}$ is the average life period of the vectors and although it depends implicitly on time, the variation of the average temperature is very slow. Therefore, using the adiabatic theory, the latter term is equal to zero, an expected result if we consider that $\mu_V(T) ^{-1}$ is the average life period of the mosquitoes. So we finally get: 
\begin{eqnarray*}
EV(t;T) &=&  \lambda(t;T)  V_E(t) +\int_0^t {   \lambda(t;T) G(t;T) U(t-s;T) ds }   \nonumber \\ 
&-&   \lambda(t;T)  \int_0^t {   G(t;T) U(t-s;T) ds }
\end{eqnarray*}

%%%%%%%%%%%%%%%%%%%%%%%%%%%%%%%%%%%%%%%%%%%%%%%%%%%%%%%
%%%%%%%%%%%%%%%%%%%%%%%%%%%%%%%%%%%%%%%%%%%%%%%%%%%%%%%
%%%%%%%%%%%%%%%%%%%%%%%%%%%%%%%%%%%%%%%%%%%%%%%%%%%%%%%

\subsection{The reproductive number $R_0$}
\label{derivationRzero}
The probability densities for the mean lifetime and the latent period of vectors are $f_\mu$ and $f_\sigma$ respectively and the corresponding survival functions are $e^{-\mu_V(T) t}$ and $\bar{F}(t;T)$.

In general (not only for exponentials) a vector in a latent state, leaves that state because:
\begin{itemize}
\item It becomes infectious before it dies.
\item It dies before it becomes infectious.
\end{itemize}
Considering the following random variables: 
\begin{itemize}
\item $t_E$: the latency time, i. e. the time the vector leaves the latency state.
\item $t_\mu$: the mortality time, i. e. the time at which the vector dies before leaving the latency state.
\end{itemize}
With the average values of $<t_E> = \tau_E$ and $<t_\mu> = \tau_\mu$, we can define $\mu_V = {\tau_{V_\mu}}^{-1}$ and $\tau_{E_V} = \sigma_{V}^{-1}$.

Using an exponential distribution for the mean lifetime, we now consider the random variable $t_i,$ which is the time that a vector remains in a latent state (before becoming infectious or dying), then:
\begin{equation}
t_i = f_\sigma (t;T) e^{-\mu_V t} + \mu_V e^{-\mu_V t} \bar{F}(t;T) 
\end{equation}

The average time the vector remains in the latency state is:
\begin{equation}
\tau_{VE} = \int_0^{\infty} t \Bigg[  f_\sigma (t;T)  + \mu_V  \bar{F}(t;T) \Bigg] e^{-\mu_V t} dt
\end{equation}
Using a Gamma distribution for the mean latency period, with integer shape parameter, and solving the last integral we obtain:

\begin{equation}
\tau_{VE} = \frac{1}{\mu} \Bigg[  1 - \Bigg( \frac{ k \sigma_V }{ \mu_V + k \sigma_V} \Bigg)^k \Bigg]
\end{equation} 
And using the following definition of $R_0$ for the SEI (and SEIR) model:
\begin{eqnarray*}
R_0 &=&  \left( 
\begin{array}{c}
\mathrm{Number~of} \\
\mathrm{contacts} \\
\mathrm{per~time}
\end{array}
\right)
\times
\left( 
\begin{array}{c}
\mathrm{Probability~of} \\
\mathrm{transmission} \\
\mathrm{per~contact}
\end{array}
\right)\\
&\times&
\left( 
\begin{array}{c}
\mathrm{Duration} \\
\mathrm{of} \\
\mathrm{infections}
\end{array}
\right)
\times
\left( 
\begin{array}{c}
\mathrm{Probability~of} \\
\mathrm{surviving} \\
\mathrm{exposed~stage}
\end{array}
\right)
\end{eqnarray*}
To calculate the $R_0$ for a vector-host model, we need to make the product of the basic reproductive numbers for the vectors and for the hosts, so we get:
\begin{eqnarray*}
R_0 &=& \beta_r \theta(T) b_H \beta_r \theta(T) b_V \frac{1}{\mu_V} \frac{1}{\gamma} \Bigg( \frac{1}{\mu_V} - \frac{1}{\mu_V} \Bigg[  1 \\
&-& \Bigg( \frac{ k \sigma_V }{ \mu_V + k \sigma_V} \Bigg)^k \Bigg] \Bigg) \frac{V(t)}{H}
\end{eqnarray*}
By operating each of the terms we obtain:
\begin{equation}
R_0 = b^2 \ \theta(T) ^2 \ \frac{p_H p_V }{\mu_V \ \gamma}  \left( \frac{k \ \sigma_V  }{\mu_V(T) +  k \ \sigma_V } \right)^k  \frac{V(t)}{H}
\end{equation}

%%%%%%%%%%%%%%%%%%%%%%%%%%%%%%%%%%%%%%%%%%%%%%%%%%%%%%%
%%%%%%%%%%%%%%%%%%%%%%%%%%%%%%%%%%%%%%%%%%%%%%%%%%%%%%%
%%%%%%%%%%%%%%%%%%%%%%%%%%%%%%%%%%%%%%%%%%%%%%%%%%%%%%%
%\newpage
%\bibliography{myreferences}

%\end{document}

\end{document}